\documentclass[twocolumn]{aastex631}
\turnoffedit
\usepackage{amsmath}
\usepackage{multirow}
\usepackage{verbatim}

\shorttitle{AASTeX v6.3.1 Sample article}
\shortauthors{Francis et al.}
\graphicspath{{./}{figures/}}

\begin{document}


\title{Gap Opening and Inner Disk Structure in the Strongly Accreting Transition Disk of DM Tau}

\author[0000-0001-8822-6327]{Logan Francis}
\affiliation{Department of Physics and Astronomy, University of Victoria \\
3800 Finnerty Road, Elliot Building \\
Victoria, BC, V8P 5C2, Canada}
\affiliation{loganfrancis3@uvic.ca}

\author[0000-0003-2458-9756]{Nienke van der Marel}
\affiliation{Leiden Observatory, Leiden University, P.O. Box 9513, 2300-RA Leiden, The Netherlands}
\affiliation{Department of Physics and Astronomy, University of Victoria \\
3800 Finnerty Road, Elliot Building \\
Victoria, BC, V8P 5C2, Canada}

\author[0000-0002-6773-459X]{Doug Johnstone}
\affiliation{NRC Herzberg Astronomy and Astrophysics, 5071 West Saanich Rd, Victoria, BC, V9E 2E7, Canada}
\affiliation{Department of Physics and Astronomy, University of Victoria \\
3800 Finnerty Road, Elliot Building \\
Victoria, BC, V8P 5C2, Canada}

\author{Eiji Akiyama}
\affiliation{Division of Fundamental Education and Liberal Arts, Department of Engineering,
Niigata Institute of Technology
1719 Fujihashi, Kashiwazaki, Niigata 945-1195, Japan}

\author{Simon Bruderer}
\affiliation{Max-Planck-Institut für Extraterrestrische Physik, Giessenbachstrasse 1, 85748 Garching, Germany}

\author[0000-0001-9290-7846]{Ruobing Dong}
\affiliation{Department of Physics and Astronomy, University of Victoria \\
3800 Finnerty Road, Elliot Building \\
Victoria, BC, V8P 5C2, Canada}

\author{Jun Hashimoto}
\affiliation{Astrobiology Center, National Institutes of Natural Sciences, 2-21-1 Osawa, Mitaka, Tokyo 181-8588, Japan}
\affiliation{Subaru Telescope, National Astronomical Observatory of Japan, Mitaka, Tokyo 181-8588, Japan}
\affiliation{Department of Astronomy, School of Science, Graduate University for Advanced Studies (SOKENDAI), Mitaka, Tokyo 181-8588, Japan}

\author[0000-0003-2300-2626]{Hauyu Baobab Liu}
\affiliation{Institute of Astronomy and Astrophysics, Academia Sinica, 11F of Astronomy-Mathematics Building, AS/NTU No.1, Sec. 4,
Roosevelt Rd, Taipei 10617, Taiwan, ROC} 

\author{Takayuki Muto}
\affiliation{Division of Liberal Arts, Kogakuin University, 1-24-2 Nishi-Shinjuku, Shinjuku-ku, Tokyo 163-8677, Japan}

\author[0000-0002-9024-4150]{Yi Yang}
\affiliation{National Astronomical Observatory of Japan (NAOJ), National Institutes of Natural Sciences (NINS), 2-21-1 Osawa, Mitaka, Tokyo 181-8588,Japan}
\affiliation{Department of Astronomy, The University of Tokyo, 7-3-1 Hongo, Bunkyo-ku, Tokyo 113-0033, Japan}



\begin{abstract}
Large inner dust gaps in transition disks are frequently posited as evidence of giant planets sculpting gas and dust in the disk, or the opening of a gap by photoevaporative winds. Although the former hypothesis is strongly supported by the observations of planets and deep depletions in gas within the gap some disks, many T Tauri stars hosting transition disks accrete at rates typical for an undepleted disk, raising the question of how gap opening occurs in these objects. We thus present an analysis of the structure of the transition disk around the T Tauri star DM Tau, which is strongly accreting ($\sim 10^{-8.3}~\text{M}_\odot~ \text{yr}^{-1}$) and turbulent ($\alpha=0.078 \pm 0.02$). Using the DALI thermochemical code, we fit disk models to simultaneously reproduce the accretion rate, high level of turbulence, the gas traced by ALMA band 6 observations of $^{12}$CO, $^{13}$CO, and C$^{18}$O J=2--1 lines, and the observed dust emission from the mm continuum and spectral energy distribution. We find a shallow depletion in gas surface density of $\sim 10$ relative to the outer disk and a gas rich inner disk is consistent with the observations. The planet mass of $<1$ M$_\text{Jup}$ implied by the gap depth is in tension with predictions for dust trapping in a highly viscous disk, which requires a more massive planet of of $\sim10$M$_\text{Jup}$. Photoevaporative models including a dead zone can qualitatively reproduce some features of the DM Tau disk, but still struggle to explain the high accretion rates and the observed mm continuum flux.

\end{abstract}

\keywords{{stars: formation - protoplanetary disks - planetary systems – protoplanetary disks - accretion, accretion disks - stars: variables: T Tauri, Herbig Ae/Be}}


\section{Introduction} \label{sec:intro}

Protoplanetary disks with a cavity in their dust distributions --- transition disks --- are a key topic of research, as their observed morphology has been variously interpreted as evidence of ongoing planet formation, disk dispersal by photoevaporation, and dead zones or grain growth in the inner disk. 
Historically, the dust rings have been identified from their spectral energy distribution by a deficit of infrared emission relative to that expected for a smooth disk, implying the absence or depletion of warmer dust close to the star \citep{Espaillat2014}. With the advent of the Atacama Large Millimeter/Submillimeter Array (ALMA), dust rings at a few 10s of au are now routinely detected and resolved at mm/sub-mm wavelengths \citep[e.g.][]{Francis2020}. Similarly, high-resolution ALMA observations have confirmed the presence of inner dust disks at a few au scales in many transition disk cavities \citep{Pinilla2019,Pinilla2021,Francis2020} that were previously identified in SED studies through the presence of excess near-infrared emission, which implied a hot inner dust disk existed close to the star \citep{Espaillat2007}, although the amount of mm-dust and NIR excess is not found to be correlated \citep{Francis2020}.

A large number of  transition disks are also actively accreting, implying the presence of gas within the dust cavity and/or a gas-rich inner disk. Demographic studies of transition disk hosting stars in several star forming regions \citep[Taurus, Ophiuchus, and Lupus,][]{Najita2007,Najita2015,Manara2016}
indicate that their median accretion rates are lower (by e.g. typically $\sim 1$ dex in Taurus) 
than non-transition T Tauris with the same disk mass. 
In a survey of transition disks with large dust gaps ($\gtrsim 20$ au) and high accretion rates ($\dot{M} \gtrsim 10^{-9}~M_\odot~\text{yr}^{-1}$) , \cite{Manara2014} found that  $\sim 80\%$ of the sample had accretion rates comparable to classical T Tauri stars, independent of the size of their dust cavities, suggesting that however the dust cavity in these disks is created must also allow for significant levels of accretion onto the star to continue. \emph{Whether the strongly accreting stars hosting transition disks are thus a distinct class of object and what mechanism(s) may be responsible for opening their dust gaps remains unclear.} 

A long-standing alternative to the planet hypothesis for the origin of transition disk gaps is inside-out dispersal of a protoplanetary disk by photoevaporation during the final stages of its evolution \citep{Shu1993}. However, these models have struggled to explain transition disks with high accretion rates and over-predict the number of non-accreting ``relic'' disks \citep{Owen2010,Owen2011,Picogna2019} whereas planet-disk interaction models allow for both gap opening and a high stellar accretion rate \cite{Lubow2006}. A dead zone in the inner disk can also produce a dust trapping pressure bump at the transition to the ionized outer disk where the magneto rotational instability is active \citep{Regaly2012,Pinilla2016}, but these models have not shown what this implies for the stellar accretion rate. Recent modelling work by \cite{Garate2021} combines X-ray photoevaporation and a dead zone in the inner disk with a lower viscous $\alpha$ parameter,
and shows that a gas depleted gap can be quickly opened ($< 1 $ Myr) in the disk, producing a dispersing outer disk and a separated gas rich inner disk which supplies accretion onto the star and is long-lived due to the longer viscous timescale. Including dust evolution in their model results in a large ($10-20$ au) outer dust ring, and a compact inner dust disk.  This model is thus able to realistically explain a much larger proportion of the observed transition disk population, though the exact population produced is sensitive to the treatment of the $\alpha$-viscosity parameterization of the dead zone. 

Given the distinct predictions made for each of the observable components (i.e., the SED, mm dust, gas, and accretion rate) of a transition disk by different gap opening models, a holistic modelling approach considering all components for a given disk is a powerful means of distinguishing between them. The gas structure of the disk within the dust cavity is particularly constraining, as deeper gaps are generally expected from photoevaporation than gap opening by planets.  In this work, we thus present a case study of DM Tau, a strongly accreting T Tauri star hosting a transition disk. We use the DALI (Dust And LInes) thermochemical code \citep{Bruderer2013} to interpret high resolution ALMA observations of DM Tau in the mm dust continuum and multiple CO lines, in combination with the disk SED and accretion rate measurements, with the goal of constraining the gas and dust structure within the cavity and evaluating the likelihood of different gap opening mechanisms.

The remainder of this paper is organized as follows: in Section \ref{sec:dm_tau}, we summarize the structure of the DM Tau transition disk known from previous studies, while in Section \ref{sec:obs}, we describe the observations of DM Tau used in this paper. In Section \ref{sec:modelling} we describe the DALI code, the parameters of the model used for DM Tau, and our model fitting approach. In \ref{sec:results}, we outline properties of the fiducial DALI model for DM Tau, discuss how well constrained key parameters are by the observations, and outline the effects of changing these parameters on the DALI model. In Section \ref{sec:disc}, we discuss our results in the context of different models for gap opening, and we conclude with a summary of our findings in Section \ref{sec:conc}.

\section{The DM Tau Transition Disk}
\label{sec:dm_tau}

DM Tau is a well-studied 0.39 M$_\odot$ and 0.24 $L_\sun$  T Tauri star of spectral type M1 ($T_\text{eff}\sim 3580$ K) with a low level of extinction ($A_V$ = 0.5) \citep{Francis2020} located 144 pc away from the Sun \citep{GaiaCollaboration2021} in the Taurus molecular cloud. Early Spitzer SED studies identified a transition disk around DM Tau with a dust cavity radius of $\sim 3$ au \citep{Calvet2005}, whereas SMA observations of the disk showed a larger cavity at $\sim 20$ au \citep{Andrews2011}. Followup high resolution dust continuum observations with ALMA showed that there is in fact both an outer ring at $\sim 21.00 \pm 0.02$ au and an inner ring at $3.16^{+0.22}_{-0.23}$ au \citep{Kudo2018}.
Asymmetries in the dust continuum emission of the 21 au ring with moderate contrast $\le 20\%$ have also been identified, which could be dust-trapping anticyclonic vortices triggered by planets in the gap \citep{Hashimoto2021}. 

Previous searches for planets towards DM Tau have not resulted in a robust detection. A Keck/NIRC2 sparse aperture masking (SAM) search identified a K band point source at 6 au at $11.0 \pm 0.3$  mag  with 4.27 $\sigma$ significance \citep{Willson2016}, however, emission from the inner disk can be spuriously identified as planets in SAM observations \citep[][]{Currie2019,Blakely2022}. \cite{Uyama2017} observed DM Tau in the Strategic Exploration of Exoplanets and Disks with Subaru (SEEDS) survey with Subaru, and did not detect any companions down to an H-band $5\sigma$ contrast limit at 0.25\arcsec of 14.76 mag. We discuss potential upper limits on the masses of planets around DM Tau from these observations in Section \ref{ssec:planets}.

DM Tau is particularly notable for its active accretion at a level of $\sim 10^{-8.3}~\text{M}_\odot~ \text{yr}^{-1}$, which is comparable to other T Tauri stars hosting non-transition protoplanetary disks \citep{Manara2014}. Furthermore, DM Tau is one of only a handful of disks where turbulence has been detected through broadening of the $^{12}$CO line \citep{Flaherty2020}, indirect evidence of high turbulence has also been found through SED studies of DM Tau which suggest the vertical distribution of dust grains is less settled towards the midplane than for other disks \citep{Qi2019}. The level of turbulence in protoplanetary disks is generally described by the $\alpha$ parameter, which relates an assumed turbulent viscosity $\nu$ to the local disk scale height $H$ and sound speed $c_s$ by $\nu=\alpha H c_s$ \citep{Shakura1973,Pringle1981}. Typical values of $\alpha$ inferred from observations are $\alpha = 10^{-4} - 10^{-3}$, indicating a very low level of turbulence, indeed, in many disks values of $\alpha$ from turbulent broadening of emission lines are only an upper limit \citep{Flaherty2015,Flaherty2018,Flaherty2020,Dullemond2018}. The broadening of the $^{12}$CO line of DM Tau observed by \cite{Flaherty2020} can be translated to a much higher $\alpha$ value of $\alpha=0.078 \pm 0.02$. As the disk accretion rate is proportional to the local viscosity and gas surface density, the high $\alpha$ value of DM Tau may thus be linked to the strong accretion towards the star. Furthermore, when combined with an estimate of the accretion rate, the measurement of $\alpha$ allows an independent estimate of the gas surface density near the star.

\section{Observations}
\label{sec:obs}

The observations of DM Tau used in this work consist of long-baseline Band 6 ALMA observations covering the thermal dust continuum and the $^{12}$CO (230.538 GHz), $^{13}$CO (220.39868420 GHz), and C$^{18}$O (219.56035410 GHz) J=2--1 lines, in combination with the spectral energy distribution (SED), stellar properties, and the accretion rate derived from VLT/X-Shooter observations. In this Section, we provide the details of the ALMA data reduction and imaging, and outline the collection of the other observational data from the literature. 

\subsection{ALMA observations}

\begin{figure*}[htb]
    \centering
    \includegraphics[]{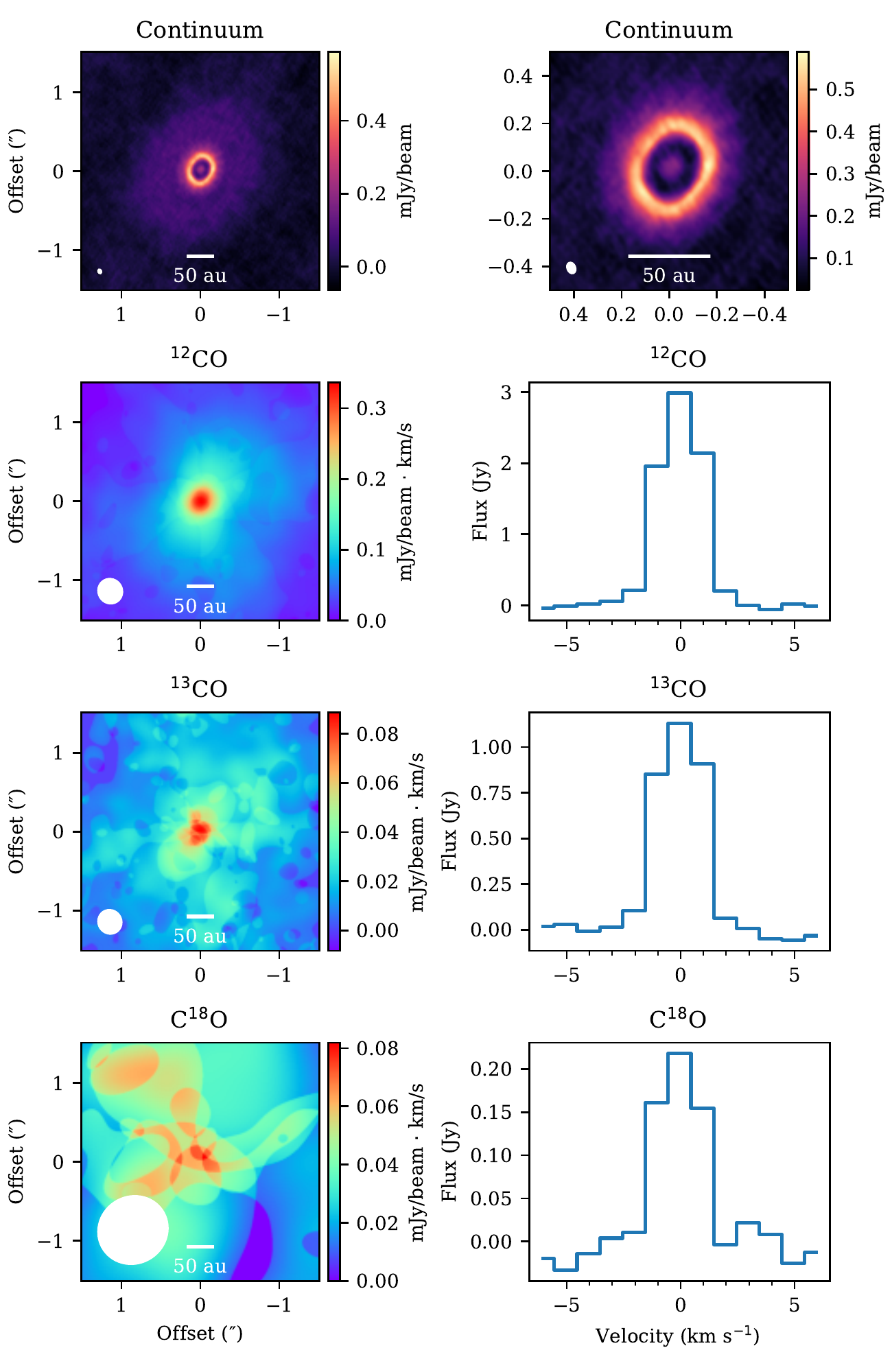}
    \caption{Top panels: 1.3 mm dust continuum images of DM Tau and zoom in on the 21 au dust cavity; lower left panels: moment 0 maps of each observed CO line; lower right panels: Flux of each CO line integrated over the disk vs the velocity. The beam for the continuum and moment 0 images is shown in the bottom-left of each panel as a white ellipse.}
    \label{fig:alma_gallery}
\end{figure*}

Data from two ALMA projects, 2013.1.00498.S and 2017.1.01460.S were collected from the ALMA archive; details on the observing setup and spectral window configuration for each project are provided in Table \ref{tab:alma_projects}. The observations of DM Tau have been previously published in \cite{Kudo2018}. The data was calibrated manually by ALMA staff for 2013.1.00498.S and with the ALMA pipeline for 2017.1.01460.S. Imaging of the calibrated data was performed in \texttt{CASA} version 5.6.1-8 with the \texttt{tclean} task using a pixel scale of 5 milliarcsec, briggs robust weighting of the data, and multiscale cleaning for the continuum image. Self-calibration was not applied as the S/N on the longest baselines tracing disk scales of interest is too low to obtain an impactful improvement in image quality. The \texttt{tclean} parameters used are summarized in Table \ref{tab:images} and the imaging procedure was as follows: a continuum image was created using the broadband spectral windows to provide a total continuum bandwidth of 4 GHz at a reference frequency of 224.805 GHz, while a robust value of 0.5 was employed for a balanced tradeoff between resolution and S/N, and multiscale cleaning \citep{Cornwell2008} was used to better capture the extended outer disk emission. To create image cubes for each CO line, the continuum was first subtracted in the $uv$-plane using the \texttt{CASA} \texttt{uvcontsub} task before imaging with \texttt{tclean} using a channel width of 1 km/s. For optimum sensitivity to extended line emission, moderate uv-tapering and a robust value of 2.0 was used during imaging. Multiscale cleaning was not employed for the DM Tau line emission because the tapered beam size is relatively large compared to the scale of the CO emission. Moment 0 maps were produced from the line cubes by collapsing the line cubes using the quadratic centroiding method of the \texttt{bettermomments} software\citep{Teague2018}. The final continuum and moment 0 maps and the line profiles integrated over the disk are displayed in figure \ref{fig:alma_gallery}.

Both the previously known inner and outer ring of DM Tau \citep{Kudo2018,Francis2020} are detected, and bright but poorly resolved emission from the $^{12}$CO and $^{13}$CO isotopologues is detected within the 21 au dust cavity of DM Tau. The C$^{18}$O emission is only marginally detected after significant $uv$-tapering of the data and is thus totally  unresolved at the scale of the outer dust ring. 

\begin{deluxetable*}{lcc}
\tablecaption{ALMA Projects}
\label{tab:alma_projects}
    \tablecolumns{2}
    \tablecaption{ALMA Project Details}
    \tablehead{\colhead{Parameter} & \colhead{2013.1.00498.S} & \colhead{2017.1.01460.S}}
    \startdata
    CASA pipeline version                       &  4.4.0\tablenotemark{a}      & 5.1.1-5 r40896 \\
    Number of Antennas                          &  44                          & 49             \\ 
    Integration time (s)                        & 846                          & 3919           \\
    Baseline range (m)                          &  15-1574                     & 135-14851      \\
    cont. window 1\tablenotemark{b}              & 128 $\times$  15265    ~@~ 216995.9192 & 128 $\times$  15265   ~@~ 232606.1459 \\
    cont. window 2                               & 128 $\times$  15265    ~@~ 232370.9192 & 128 $\times$  15265   ~@~ 217023.1459 \\ 
    $^{12}$CO window                            & 1920 $\times$  244.141 ~@~ 230736.2530 & 1920 $\times$  61.035 ~@~ 230537.5987 \\
    $^{13}$CO window                            & 960 $\times$  488.281  ~@~ 220204.2095 & 960 $\times$  122.070 ~@~ 219559.9088 \\
    C$^{18}$O window                            & 960 $\times$  488.281  ~@~ 219735.4595 & 960 $\times$  122.070 ~@~ 220398.1657 \\
    \enddata
\tablecomments{
\tablenotetext{a}{Manual calibration was performed by ALMA staff.}
\tablenotetext{b}{The spectral window setup for each project is specified in the format number of channels  x channel width (kHz) ~@~ window center frequency (MHz).}
}
\end{deluxetable*}

\begin{deluxetable*}{ccccccc}
\tablecaption{Image properties}
\label{tab:images}
\tablehead{ \colhead{Line} & \colhead{Beam FWHM\tablenotemark{a} (\arcsec)} & \colhead{Approx. Beam FWHM (au)} & \colhead{RMS} & 
\colhead{robust} & \colhead{multiscale (pix)\tablenotemark{a}} & \colhead{uvtaper (\arcsec)}}
\startdata
 cont.     & 0.045 $\times$ 0.032 & 7   & 0.018 mJy/beam             & 0.5 & 0, 6, 20, 40             & -   \\
$^{12}$CO  & 0.312 $\times$ 0.296 & 62  & 2.9 mJy/beam $\cdot$ km/s  & 2.0 & -                        & 0.3 \\
$^{13}$CO & 0.301 $\times$ 0.286  & 60  & 2.7 mJy/beam $\cdot$ km/s  & 2.0 & -                        & 0.3 \\
C$^{18}$O & 0.892 $\times$ 0.840  & 178 & 6.2 mJy/beam $\cdot$ km/s  & 2.0 & -                        & 0.3 \\
\enddata
\tablecomments{
\tablenotetext{a}{The size given is the full width at half maximum (FWHM) of the major and minor axes of the clean beam ellipse.}
\tablenotetext{b}{The pixel size is 5 milliarcsec for all images. 
}
}
\end{deluxetable*}

\begin{figure}[htb]
    \centering
    \includegraphics[]{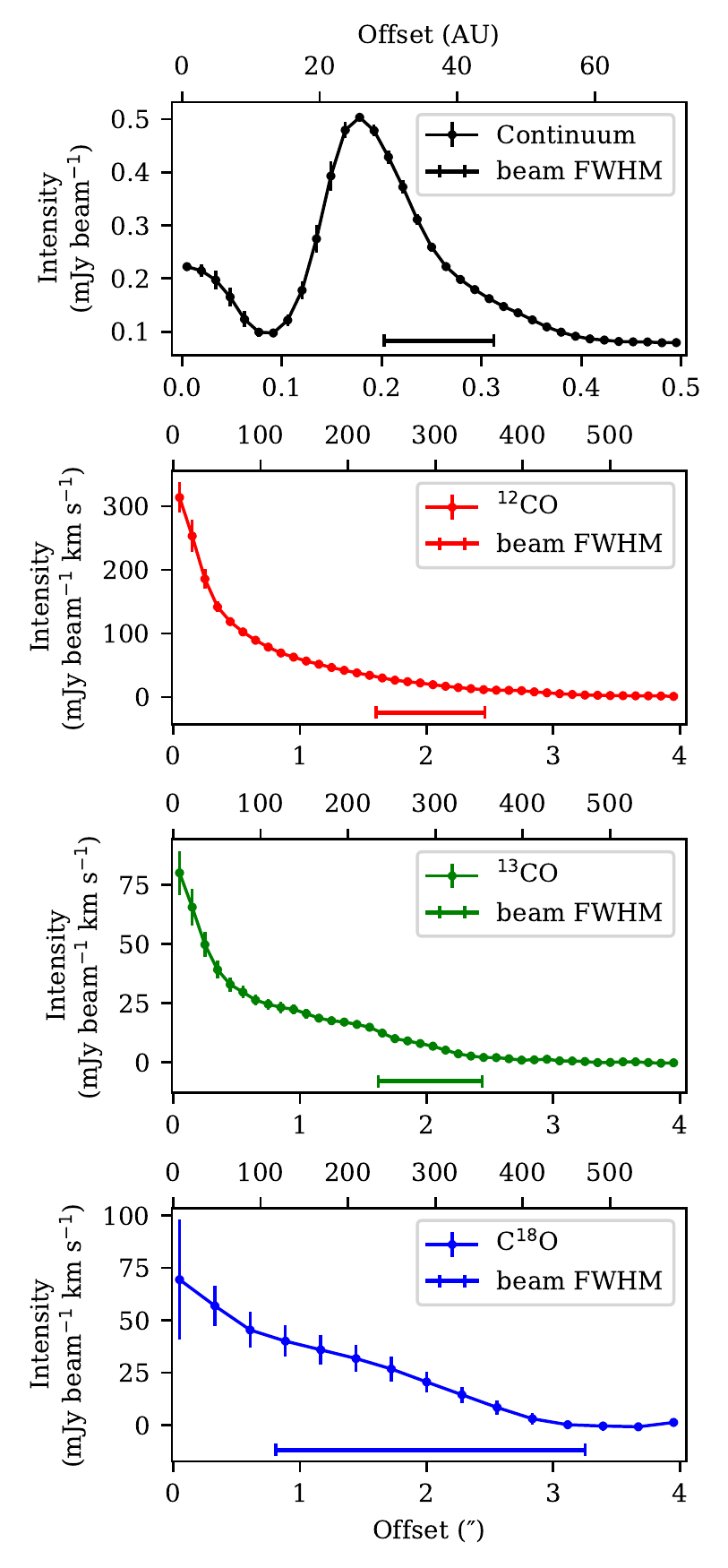}
    \caption{Deprojected and azimuthally averaged intensity profiles of the 1.3 mm dust continuum and moment 0 maps for each observed CO line. The full width at half maximum (FWHM) of the beam is shown by the scalebar.}
    \label{fig:obs_radial_profiles}
\end{figure}

To further quantify the radial structure of the DM Tau disk, we deproject the continuum image and moment maps and compute the azimuthally averaged intensity in annular bins. The disk center used for the deprojection is 04:33:48.749, +18.10.09.640 (ICRS J2000) while the inclination and position angle are $i=35.2 \pm 0.7^\circ$ and $\theta=157.8 \pm 1.0^\circ$, respectively, which were determined from fitting of the continuum visibilities by \cite{Kudo2018}. The deprojection is performed by rotating the disk images clockwise by $90^\circ - \theta$, then interpolating the image to stretch the declination axis by a factor of $1/\cos(i)$. This process does not conserve flux, as would be required for accurate deprojection of optically thin emission, however, the difference in flux is small ($\sim 20\%)$, and it is sufficient for our purposes of examining the radial structure and comparison with the DALI models, which are deprojected by the same procedure (Section \ref{ssec:raytracing_models}). The resulting deprojected and azimuthally averaged intensity profiles are shown in Figure \ref{fig:obs_radial_profiles}. The radial peak of the dust continuum emission from the outer disk is at $\sim$ 26 au while the inner dust disk has a radius of $7.5 \pm 0.3$ au as determined by the full width at half maximum (FWHM) of a Gaussian fit \citep{Francis2020}. In contrast to other transition disks observed in CO \citep{vanderMarel2016b,vanderMarel2021}, no gap is visible in the CO emission from all isotopologues, however we note that the 21 au dust cavity is not well resolved ($^{12}$CO, $^{13}$CO) or completely unresolved (C$^{18}$O) by the larger beam in the CO observations.

\subsection{Spectral Energy Distributions}
\begin{figure}[htb]
    \centering
    \includegraphics[]{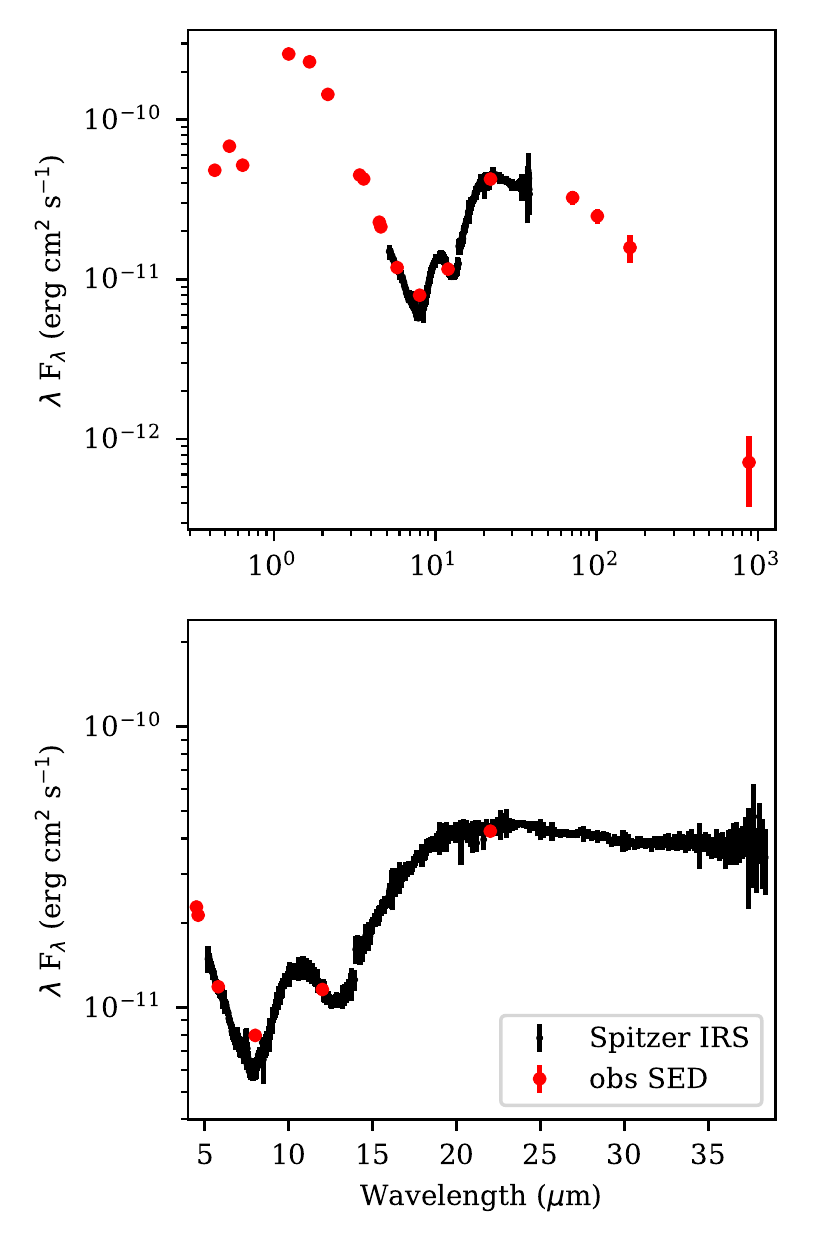}
    \caption{Spectral energy distribution for DM Tau, red circles are photometric measurements, black dots are Spitzer IRS spectra. Top panel: overall SED; Bottom panels: zoom in on the near to mid infrared portion.}
    \label{fig:obs_seds}
\end{figure}

The SED of DM Tau was constructed using the data from \citet{Francis2020}, which consists of optical to mid-infrared photometry from \textit{BVRI} surveys, \textit{2MASS (JHK)}, \textit{WISE}, and \textit{Spitzer-IRAC}. Additionally, near to mid infrared spectra  were retrieved from the Cornell Atlas of \textit{Spitzer}/IRS Sources (CASSIS) \citep{Lebouteiller2011}. The resulting SED is shown in Figure \ref{fig:obs_seds} with no dereddening correction applied. A prominent dip in the overall near-infrared emission is present, indicating a large dust cavity, but some emission around the 10 $\mu$m Si feature is also detected, implying that hot dust exists close to the star.

\subsection{Stellar Properties}
\label{ssec:stellar}

The stellar properties of DM Tau (see Section \ref{sec:dm_tau})
were derived from fits of stellar models to the SED as described in Section 2.3 of \citet{Francis2020}, except for the accretion rate, which was measured using UV to optical spectra from VLT/X-SHOOTER observations by \citet{Manara2014}. While the stellar properties determined by \cite{Francis2020} have been corrected for GAIA distance measurements (144 pc), the accretion rates were derived with pre-GAIA distance measurements of 140 pc and the accretion rate measurement can not simply be rescaled; however, the typical uncertainty in the accretion rate of $\sim0.5$ dex is much larger source of uncertainty than the distance measurement, which does not differ significantly from the updated GAIA value. 


\section{Modelling}
\label{sec:modelling}

Resolved measurements of CO emission in disks can also be used as a tracer of the gas surface density, however,
translating the CO emission to a gas surface density map is not simple; the CO abundance varies throughout the disk due to photodissociation in the upper layers and freezeout towards the cold midplane. 
The DALI thermochemical modelling code \citep{Bruderer2012} addresses these issues by solving for the gas temperature and CO abundance in a model disk, provided an input gas and dust structure and model for the stellar radiation field. DALI has previously been used for determining the gas surface density within transition disk gaps from ALMA CO observations \citep{Bruderer2013,vanderMarel2015,vanderMarel2016b}, and moreover, also performs the dust radiative transfer necessary for comparison with the mm dust continuum and SED. 

The steps involved in running a DALI disk model are the following: starting from an input density structure and stellar SED, DALI solves the continuum radiative transfer equations to obtain the dust temperature and radiation field throughout the disk, then self-consistently solves for the gas temperature and chemical abundances. Once a solution is found, raytracing simulations are performed to produce model continuum images, line cubes, and SEDs, which can be compared with the observations. The DALI density structure is two dimensional in radius, $r$, and height above the disk midplane, $z$, with azimuthal symmetry of the disk assumed,  which is appropriate for DM Tau given the low contrast of the known asymmetries in the 21 au dust ring \citep{Hashimoto2021}. 

In the following Sections, we describe the detailed structure of the DALI models used to interpret the DM Tau observations and determine the gas and dust structure within the 21 au cavity.

\subsection{Chemical Network and Gas Composition} 
The underlying chemical network is the same as in previous DALI studies, and is based upon the reactions in the UMIST2006 database \citep{Woodall2007} with some extensions for non gas-phase reactions described by \cite{Bruderer2009} and \cite{Bruderer2012}. Molecular data for collisions and radiative rates are taken from the LAMDA database \citep{Schoier2005}; the calculation of the reaction rate coefficients for CO in particular are described by \cite{Yang2010} and \cite{Jankowski2005}.  The elemental composition of the disk gas used in the DALI chemical network is assumed to be that of the diffuse ISM \citep{Bruderer2012}, and the gas phase abundances of each element relative to hydrogen used are listed in Table \ref{tab:fitting_params_dm_tau}. Although we assume ISM abundances for carbon and oxygen, gas mass measurements of DM Tau from HD imply that the CO abundance may be lower than the ISM value, a point which we return to in our discussion of the modelling caveats in Section \ref{ssec:caveats}.
\subsection{Surface density structure}

\begin{figure}[htb]
    \centering
    \includegraphics{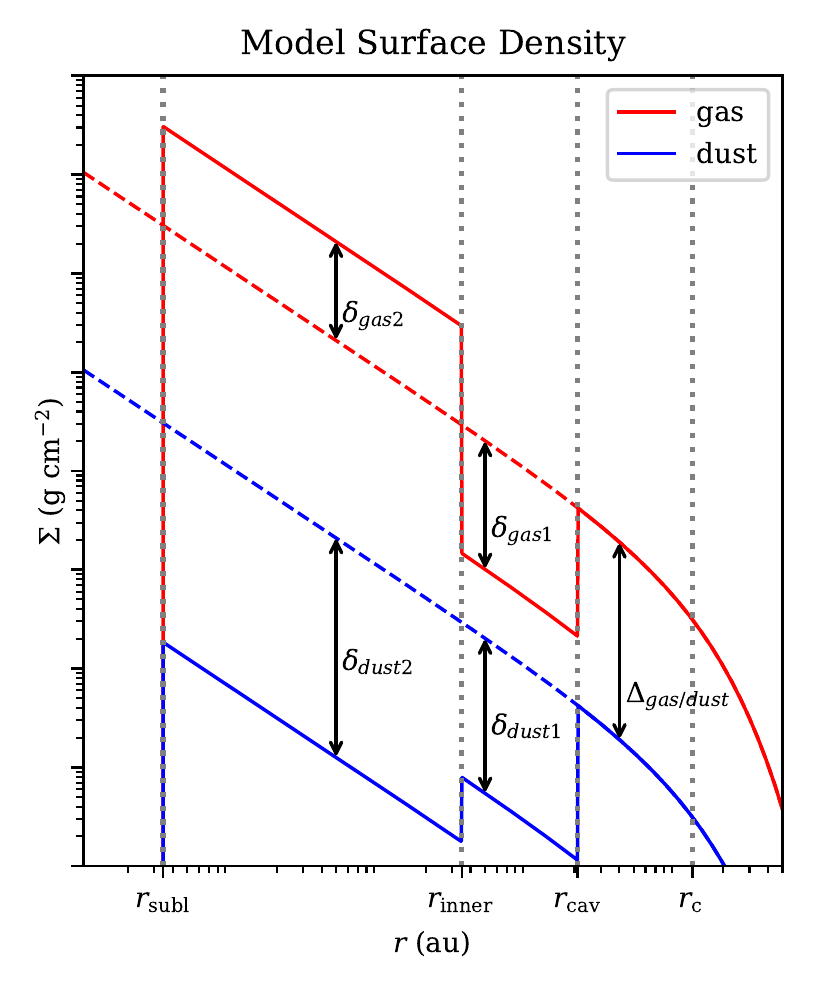}
    \caption{Surface density structure of the gas and dust model in DALI used to simulate the DM Tau transition disk. The dashed lines show the truncated power law profile for the gas and dust with no depletion.}
    \label{fig:dm_tau_surf_dens}
\end{figure}

The radial surface density structure for DM Tau is shown in Figure \ref{fig:dm_tau_surf_dens}, and closely follows the transition disk structure  previously implemented in DALI by \citet{Bruderer2013}, with some modifications to account for the inner dust ring. The surface density profile for gas in the outer disk is a radial power law which transitions to an exponential cut-off at a critical radius $r_c$, as motivated by analytical models of a self-similar viscously evolving accretion disk with a time-independent radial distribution of viscosity $\nu \propto r^\gamma$: \citep{Lynden-Bell1974,Hartmann1998}:
\begin{equation}
\Sigma(r) = \Sigma_c \left(\frac{r}{r_c}\right)^{-\gamma} {\rm exp}\left[-\left(\frac{r}{r_c}\right)^{2-\gamma}\right]
.\end{equation}
We assume $\gamma=1$, which corresponds to a vertically isothermal disk with an alpha viscosity prescription $\nu=\alpha c_s H$ \citep{Pringle1981} and radial temperature profile $T \propto r^{-1/2}$ \citep{Hartmann1998}.
The surface density for the dust in the outer disk follows the same relation, but with a constant scaling set by the gas to dust ratio $\Delta_\text{gas/dust}$. To reproduce the observed dust outer and inner dust rings of DM Tau (Section \ref{sec:dm_tau}), we introduce two depletion factors in surface density relative to the outer disk, $\delta_\text{dust1}$ and $\delta_\text{dust2}$, which begin at radii of $r_\text{cav} = 21$ and $r_\text{inner} = 3.5$ au respectively. We find that the power law rise in dust surface density of our model between 21 and 3.5 au can provide a good match to the surface brightness of the inner dust ring without the need to introduce additional model parameters for its outer edge and dust surface density, allowing for a simpler model.

For the gas surface density, corresponding depletion factors of $\delta_\text{gas1}$ and $\delta_\text{gas2}$ interior to the same radii as those for the dust are included, however, for some models we also allow $\delta_\text{gas2}$ to be a relative \emph{enhancement}, in order to explore the effects of variations in the gas content of the inner disk. Defining the inner edge of the model, both the gas and dust surface density are set to zero at the dust sublimation radius of 0.0343 au according to the relation $R_\text{subl}\sim 0.07 (L_*/L_\odot)^{1/2}$ au \citep{Dullemond2001}.

\subsection{Vertical structure and dust populations}
\label{ssec:vert_dust}
The vertical density structure of the gas disk is a Gaussian with scale height angle $h(r)=h_c(r/r_c)^\psi$, where $\psi$ is a radial power law index which allows for flaring of the disk and the physical scale height is $H \sim rh$. To approximate vertical dust settling, two populations of small (0.005 - 1 $\mu$m) and large (0.005 - 1000) $\mu$m dust grains are included; the small dust grains follow the gas with scale height angle $h$,
while the large dust grains are more concentrated towards the midplane and have a reduced scale height angle of $\chi h$, with $\chi$ fixed to 0.2. The dust surface density is partitioned between the two populations by a large grain fraction $f_{ls}$ and (1 - $f_{ls}$) for the small grains. The large grain fraction is by default set to $f_{ls}$=0.85 throughout the disk. The dust composition and opacities used for the radiative transfer are a standard ISM composition \citep{Weingartner2001}, and are described in further detail by \cite{Bruderer2013}. Polycyclic Aromatic Hydrocarbons (PAHs) are also included in the model, which are largely important as a source of optical/UV opacity for heating the gas and an additional pathway for H$_2$ formation \citep{Bruderer2013}. The PAHs are assumed to be entirely in the form of neutral $\text{C}_{100}\text{H}_{25}$, and their opacities are calculated from the work of \cite{Li2001} and \cite{Draine2007a} using the methods of \cite{Visser2007}. The PAHs are assumed to be destroyed in regions of the disk where the local UV flux is stronger than $10^6$ times the interstellar radiation field. The abundance of PAHs throughout the disk is set by the parameter $f_\text{PAH}$, which controls the fraction of PAHs in the disk with respect to an ISM PAH fraction of 5\% of the dust mass \citep{Draine2007b}.

\subsection{Stellar model and radiation field}

The central star is approximated as a blackbody with temperature 3580 K and luminosity 0.24 $L_\sun$ based on the stellar properties of DM Tau (Section \ref{sec:dm_tau}). UV heating from accretion is included by adding the contribution of an additional blackbody with $T_\text{acc} = 10^4$ K and accretion luminosity
\begin{equation}
\label{eqn:lacc}
L_{\rm acc} = \pi \frac{GM_* \dot{M}}{R_*},
\end{equation}
where $G$ is the gravitational constant, $M_*$ and $R_*$ the mass and radius of the star, and $\dot{M}$ the stellar accretion rate. Modelling the UV heating is particularly important for the disk chemistry, but is subject to some uncertainty due to the paucity of high quality UV observations, and different approaches using detailed stellar models have been used in other astrochemical codes \citep{Woitke2016}.  For the observed accretion rate of DM Tau of $\sim 10^{-8.3}~\text{M}_\odot~ \text{yr}^{-1}$, the accretion luminosity as a fraction of that from the star is $L_{\rm acc}/L_* = 0.24$.  This relationship assumes that all of the gravitational energy liberated by accretion is radiated away, which is a reasonable approximation for magnetospheric accretion onto pre-main-sequence stars \citep{Hartmann2016}.

Similarly, heating of the disk from stellar X-rays is taken into account by adding an X-ray luminosity of $L_{X}=10^{30}$ erg s$^{-1}$ with a thermal spectrum at T$_X=10^7$K, which is reasonably consistent with the observed X-ray luminosity of $L_{X}= 1-3 \times 10^{30}$ erg s$^{-1}$ \citep{Dionatos2019} given the intrinsic variability of X-ray emission. Finally, a background cosmic ray ionization rate $\zeta_{cr}$ of $5 \times 10^{17}$ s$^{-1}$ is assumed to be present throughout the disk.

\subsection{Model grid}

The grid used for the calculation consists of 200 cells in the radial direction and 80 cells in the vertical direction. To resolve the parts of the disk with sharp transitions in gas and dust surface density (i.e., $r_\text{subl}$, $r_\text{inner}$, and $r_\text{cav}$) the cell size grows exponentially starting from these radii, i.e, $\Delta_\mathrm{cell} = \Delta_\mathrm{min} 10^{i \delta}$, where $\Delta_\mathrm{min}$ = 0.01 au is the minimum radial cell width, $i$ is the radial cell index, and $\delta$ determines the cell growth rate. The 200 radial cells are distributed in this way to provide 50 cells from 0.01 to 3 au, 50 from 3.1 to 21 au, and 100 from 21 au to the edge of the calculation grid at 500 au.  Vertically, 80 cells are arranged to follow the scale height of the gas disk. 

\subsection{Raytracing and Model Images}
\label{ssec:raytracing_models}
Using the converged output of the thermochemical modelling, the DALI code performs raytracing simulations to produce model images and line cubes. To match the ALMA observations, the raytracing is performed with the model disk at an inclination of $35.2^\circ$ and distance of 144 pc (see Sections \ref{sec:dm_tau} and \ref{sec:obs}) with a pixel size of 5 milliarcsec. The model SED was created by raytracing the disk from 0.09 to 3000 $\mu$m and calculating $\lambda F_\lambda$ over the entire raytraced image. Sampling of the SED was performed at wavelengths approximately corresponding to the photometric data, except for the 5-20 $\mu$m range, where steps of 0.33 $\mu$m were used for comparison with the Spitzer/IRS spectrum. Model continuum images were created by raytracing at a wavelength of 1.3mm, followed by convolution with the observed continuum beam and deprojection by the same procedure as described for the observations in Section \ref{sec:obs}. Similarly, model line cubes were created assuming a keplerian velocity structure of the disk and raytracing in 0.5 km s$^{-1}$ steps (twice the observed line cube resolution of 1 km s$^{-1}$), and centered on the rest frequency of each CO line, then convolving with the observed beam. The model line cubes were also converted to moment 0 maps with \texttt{bettermomments} \citep{Teague2018} and deprojected in the same manner as the continuum. Model spectra for each CO line were created from the inclined line cubes by subtracting the continuum measured in line-free channels and calculating the total flux over the entire disk. We note that our treatment of the deprojection for both the models and observations does not conserve flux, however, the deprojection is performed identically for both, allowing a consistent comparison.

\subsection{Model fitting approach}
Due to the large computational cost of the DALI modelling, a formal model fitting approach (e.g. $\chi^2$ gradient descent) is impractical. Our modelling strategy is instead to obtain a \emph{fiducial} model which approximately reproduces the features of the observations: the mm dust continuum, the $^{12}$CO, $^{13}$CO, and C$^{18}$O moment 0 maps, and SED, with a particular focus on the structures within the 21 au dust cavity. Additionally, we wish to investigate if the gas surface density of the fiducial model within the cavity of the disk is commensurate with the stellar accretion rate, and to explore variations in the fiducial model parameters to understand the effect of the inner disk and cavity structure on the CO emission.

To compare the DALI models for the mm continumm and CO lines to the observations, we took the output deprojected images and moment 0 maps and produced azimuthally averaged radial intensity profiles in the same manner as the observations in Figure \ref{fig:obs_radial_profiles}. We also compare the model SED produced by the raytracer with the observed SED over the total wavelength range traced by the photometry and the Spitzer/IRS spectrum in the near to mid infrared.

In order to obtain a fiducial model, we first ran DALI simulations using only the dust radiative transfer and raytracing of the continuum to compared with the mm images and SED. Inspection of the mm images was first used to estimate initial values of model radii $r_\text{inner}$, $r_\text{cav}$, and $r_\text{c}$, after which we varied the outer disk dust surface density via $\Sigma_\text{gas,c}$, $\Delta_\text{gas/dust}$ and the dust depletion factors $\delta_\text{dust1}$,$\delta_\text{dust2}$ until the mm continuum and SED was approximately reproduced. The near-infrared spectrum of DM Tau displays no obvious PAH features, suggesting a low abundance of PAHs, so we varied the global PAH fraction $f_\text{PAH}$ to reproduce the lack of PAH emission, ultimately removing them from the disk entirely. We note that PAH emission features are generally excited at high temperatures in the inner disk, and thus some PAHs may still be present in the outer disk where they are shielded from destruction, however, detailed modelling of the disk PAH distribution is beyond the scope of this work. After a reasonable model for the dust was obtained, we ran full DALI simulations including the gas temperature and chemistry solver and raytracing of the CO lines. To fit the gas emission in the outer disk, we varied the outer disk gas surface density $\Sigma_\text{gas,c}$ while proportionally changing $\Delta_\text{gas/dust}$ to keep the dust model unchanged and compared the resulting DALI models with the $^{12}$CO, $^{13}$CO, and C$^{18}$O moment 0 maps and integrated line profiles. Finally, we varied the gas depletion factors in the inner disk $\delta_\text{dust1}$,$\delta_\text{dust2}$ to fit the (partially-resolved) CO emission within the 21 au cavity.

\section{Results}
\label{sec:results}

In this Section, we describe the fiducial DALI model for DM Tau, explore how well the inner disk and cavity structure are constrained by observations,  and consider the general effects of variations in the model parameters on the physical conditions of the disk. 

\subsection{Fiducial model}

\begin{figure*}[htb]
    \centering
    \includegraphics{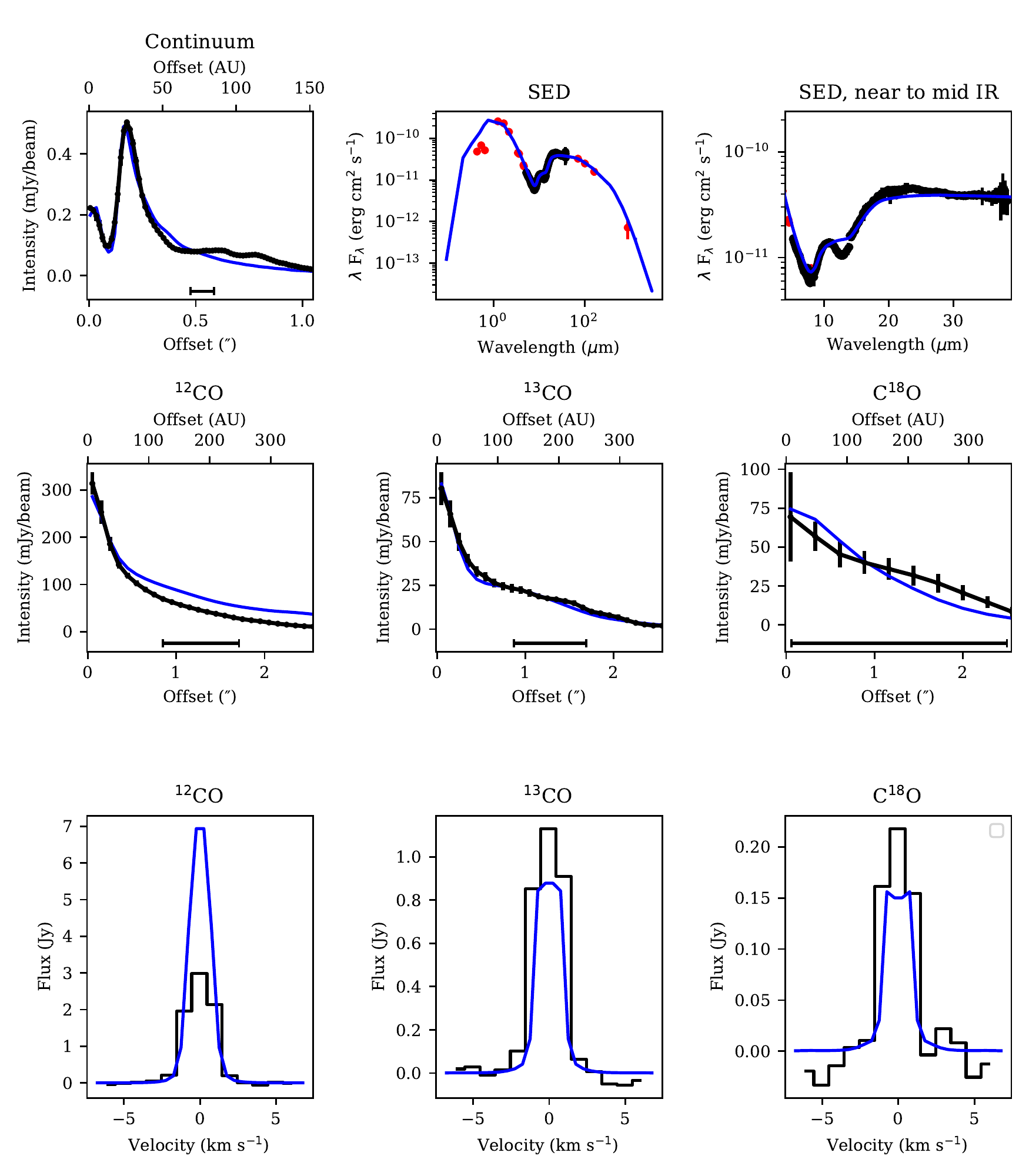}
    \caption{Comparison of the DM Tau observations (black lines and red circles for SED photometry) and fiducial model (blue lines). Top Row: deprojected and azimuthally averaged mm continuum intensity, Overall SED, and SED zoomed in on the near to mid infrared region. Middle Row: deprojected and azimuthally averaged intensity for the moment 0 maps of the $^{12}$CO, $^{13}$CO and C$^{18}$O lines. Bottom row: Flux integrated over the disk vs velocity for the $^{12}$CO, $^{13}$CO and C$^{18}$O line cubes. The spectral resolution of the observed line cubes is 1 km s$^{-1}$, while the models are sampled in 0.5 km s$^{-1}$ steps}.
    \label{fig:dm_tau_fid_radial_SED_lines}
\end{figure*}

The parameters of the fiducial DALI model fit are provided in Table \ref{tab:fitting_params_dm_tau}. To show the accuracy with which the fiducial model reproduces the observations, the output of the DALI raytracer is compared with the deprojected azimuthally averaged continuum, CO isotopologues, SED and CO line fluxes integrated over the disk in Figure \ref{fig:dm_tau_fid_radial_SED_lines}. A direct comparison of the output images and moment maps from the raytracer is provided in Figure \ref{fig:dm_tau_fid_im_res} of the appendix. 

The radial profile of the dust emission is reasonably well described by the model, however, the model underestimates the emission in the outer disk beyond $\sim 60$ au. The ALMA observations of \citet{Hashimoto2021} have shown several low contrast dust rings are present at these points in the outer disk, which we do not attempt to reproduce as our focus is on the dust cavity and inner disk. The model SED also agrees well with the observations, although the exact shape of the near-infrared silicate feature is not reproduced, specifically, the flux at $\sim 7$ microns is overestimated, suggesting some details of the inner disk structure are missing. \citet{Hashimoto2021} have performed radiative transfer modeling of the same SED and show that the dip at $\sim 7$ microns is sensitive to the dust size distribution between 1 and 3 au, and can be reproduced by a model with a depletion in the surface density of micron sized dust grains of $10^{-4}$ (their Figure 5). We have also experimented with varying the amount of small dust grains from the sublimation radius to 3.5 au and similarly find there is a small effect on the near to mid infrared shape, however, we were unable to reproduce the exact observed SED shape, which may require more careful modelling of the inner ring geometry that is beyond the scope of this paper. 

\begin{deluxetable*}{ll}
\tablecaption{DM Tau Parameter Space}
\label{tab:fitting_params_dm_tau}

\tablehead{\colhead{Parameter} &  \colhead{Value}} 
\startdata
\multicolumn{2}{c}{DM Tau}\\
$\Sigma_\mathrm{gas,c}$    & 0.65 g cm$^{-2}$\\
$\Delta_\mathrm{gas/dust}$ & 10 \\
$r_\mathrm{subl}$          & 0.0343 au \\
$r_\mathrm{inner}$         & 2.5, \textbf{3.5}, 4.5 au \\
$r_\mathrm{cav}$           & 21 au \\ 
$r_\mathrm{c}$             & 124.1  au\\
$h_c$                    & 0.03, \textbf{0.06}, 0.09 rad \\
$\psi$                   & 0.2 \\
$\delta_\mathrm{gas1}$     & 0.015, \textbf{0.15}, 1.0 \\
$\delta_\mathrm{gas2}$     & \textbf{0.15}, 1.0, 10.0  \\
$\delta_\mathrm{dust1}$    & {\color{red} 1.0(-50)}, 2.7(-2) \\
$\delta_\mathrm{dust2}$    & {\color{red} 1.0(-50)}, 6.0(-10), \textbf{6.0(-9)}, 6.0(-8), 6.0(-7), 6.0(-6) \\
$\chi$                   & 0.2 \\
$f_{ls}$ (outer disk)    & 0.85\\
L$_\odot$                  & 0.24 \\
T$_*$                      & 3580 K \\
log($\dot{M}$)           & -9.3, \textbf{-8.3}, -7.3 [log(M$_\odot$ yr$^{-1}$)] \\
L$_X$                    & $10^{30}$ erg s$^{-1}$ \\
T$_X$                    & $10^7$K \\
$\zeta_{cr}$             & $5 \times 10^{17}$ s$^{-1}$ \\
$f_\mathrm{PAH}$           & \textbf{1.0(-50)}, 1.0(-5), 1.0(-3) \\
\hline
Gas Abundance \\
\hline
H   & 1        \\
He  & 7.59(-2) \\
C   & 1.35(-4) \\
N   & 2.14(-5) \\
O   & 2.88(-4) \\
Mg  & 4.17(-6) \\
Si  & 7.94(-6) \\
S   & 1.91(-6) \\
Fe  & 4.27(-6)
\enddata
\tablecomments{Numbers written as $a(b)$ indicate $a \times 10^b$. Where a range of values is indicated, parameters denoted bold are the fiducial model parameters. Parameters in red are the non-fiducial values for the case of no inner disk. }
\end{deluxetable*}

The radial profiles of the CO isotopologue moment 0 maps in Figure \ref{fig:dm_tau_fid_radial_SED_lines} are a good match to the observed $^{13}$CO and C$^{18}$O emission, but the $^{12}$CO emission is overproduced in the outer disk. As the $^{12}$CO line is optically thick throughout the entire disk, this indicates the temperature in the outer disk of the DALI model is overestimated in the warm upper layers of the disk. The integrated line profiles in Figure \ref{fig:dm_tau_fid_radial_SED_lines} show that the $^{12}$CO flux is also overproduced by a factor of $\sim 2$, while the $^{13}$CO and C$^{18}$O flux is a reasonable match to the observations, being underestimated by $\sim 10-20\%$.

\begin{figure*}[htb]
    \centering
    \includegraphics{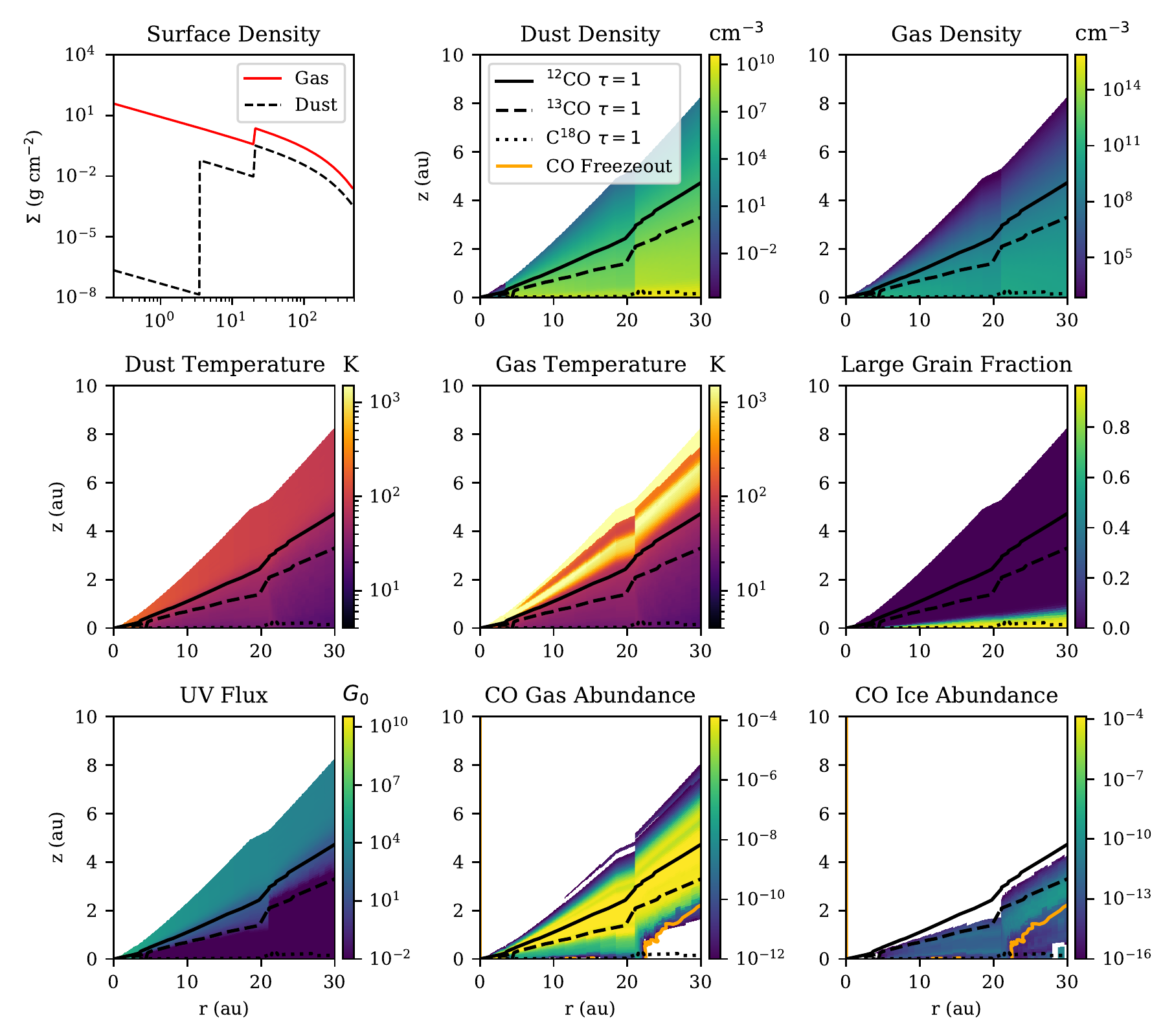}
    \caption{Surface density structure and selected 2D physical quantities from the fiducial DM Tau model plotted from 0 to 30 au. The $\tau=1$ surfaces for $^{12}$CO, $^{13}$CO and C$^{18}$O lines are overlaid on each 2D DALI panel as the solid, dashed, and dotted lines respectively while the approximate freezeout temperature of CO (25.5 K) is overlaid as the orange contour on the CO abundance panels.}
    \label{fig:dm_tau_fid_surf_dens_dali}
\end{figure*}

We now discuss the physical structure of the fiducial DALI model, and the effect of the drops in the gas and dust density at 3.5 au and 21 au on the disk temperature structure and CO abundance. In Figure \ref{fig:dm_tau_fid_surf_dens_dali}, the surface density structure of the fiducial model and the relevant physical quantities of the DALI model (gas and dust density and temperature, CO gas abundance, etc.) are shown as a function of radius and height above the midplane within 30 au. Overplotted on each panel are the $\tau=1$ surfaces for $^{12}$CO, $^{13}$CO and C$^{18}$O emission. The same panels are shown with a zoom in on the inner 5 au of the disk in Figure  \ref{fig:dm_tau_fid_surf_dens_dali_zoom}; while a figure showing the full disk is provided in the appendix (Figure \ref{fig:dm_tau_fid_dali_full_disk}). 

\begin{figure*}[htb]
    \centering
    \includegraphics{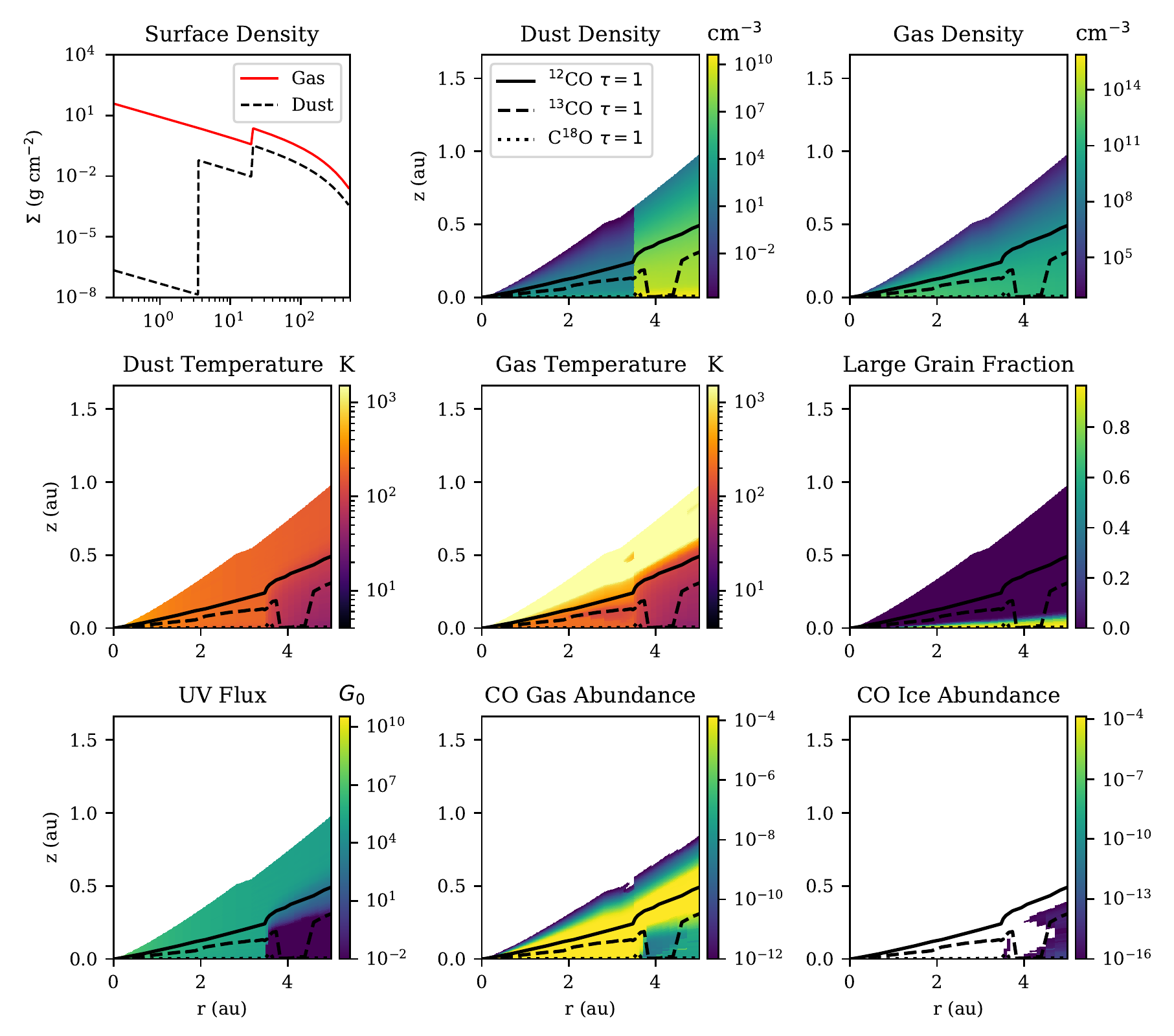}
    \caption{The same fiducial DALI model as Figure \ref{fig:dm_tau_fid_surf_dens_dali}, but zoomed in on the inner disk from 0 to 5 au.}
    \label{fig:dm_tau_fid_surf_dens_dali_zoom}
\end{figure*}

\begin{figure*}[h]
    \centering
    \includegraphics[]{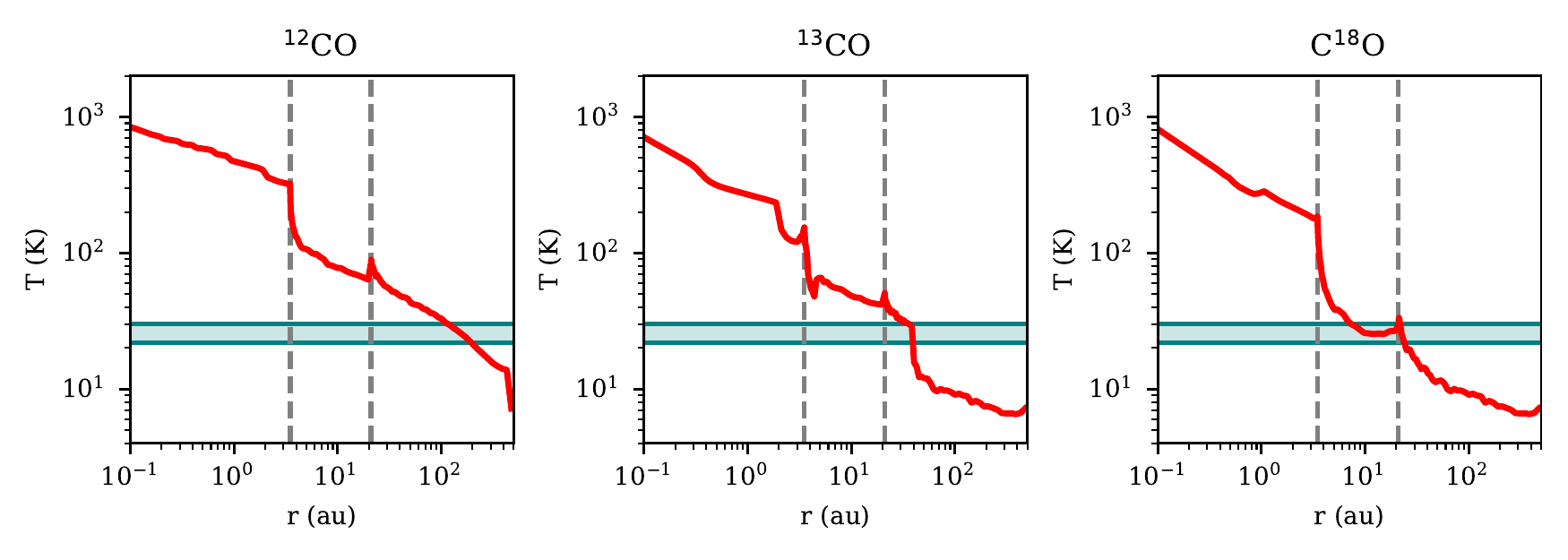}
    \caption{Gas temperature in the fiducial DALI DM Tau model at the $\tau=1$ surface for the $^{12}$CO, $^{13}$CO and C$^{18}$O emission (red lines). The dashed vertical grey lines are at the inner cavity at 3.5 au and main dust cavity at 21 au, while the shaded blue region shows the CO freezeout temperature range. }
    \label{fig:dm_tau_fid_gastemptau1}
\end{figure*}

The gas and dust temperature change significantly with height above the disk midplane due to the attenuation of the UV and X-ray flux towards deeper layers in the disk and changes in which heating mechanisms and molecular coolants are at play. Most of the gas phase CO is present in a layer of warm (few $100$ K) gas where it can survive either photodissociation in the disk atmosphere or freezeout towards the disk midplane. Moving inward radially, at the main cavity edge at 21 au the decrease in the gas and dust density has several effects. The lower dust density  causes the UV flux to penetrate deeper into the disk, especially at intermediate (z $\sim$ 1) layers in the disk, resulting in an increase in the gas and dust temperature.
The warmer and CO enriched upper layers of the disk shift downward with the decreased gas density, as do the $\tau=1$ surfaces of the CO lines. At the lower layers of the outer disk, all the CO in the gas is frozen out, but within the dust cavity the height of the CO-depleted gas drops and the CO ice abundance decreases. In Figure \ref{fig:dm_tau_fid_gastemptau1}, we show the gas temperature at each of the CO line $\tau=1$ surfaces out to 30 au, with the key radii of 3.5 and 21 au marked in dashed lines and the CO freezeout temperature indicated by the shaded blue region. While the gas and dust temperature generally increase at a given height moving into the cavity, the temperature at the $\tau=1$ surface actually drops, as the sharp drop in gas density also causes each line to become optically thick at lower and cooler regions of the disk. Within the 21 au dust cavity, both the $^{12}$CO and $^{13}$CO lines are formed in regions of the disk well above the CO freezeout temperature, on the other hand, C$^{18}$O traces cold regions of the disk midplane below the freezeout temperature until inward of $\sim 7$ au. 

Similar rapid changes also occur at the other sharp transition in the model at 3.5 au (Figure \ref{fig:dm_tau_fid_surf_dens_dali_zoom}, where the dust surface density drops by 7 orders of magnitude while the gas surface density is unchanged. The gas temperature at the $\tau=1$ surface for each line increases rapidly approaching 3.5 au as the shielding from the stellar radiation provided by the dust drops. At $\sim 4$ au, the $\tau=1$ surface of the $^{13}$CO line briefly drops to the midplane and both the CO gas and ice abundance are reduced. Here an inspection of the abundances of the other volatile species in the DALI model reveals that most of the NH$_3$ ice sublimates at this location, and much of the carbon has become incorporated in HCN instead of CO. Within 3.5 au, the UV flux towards the midplane ($z \lessapprox 0.2$) and the dust and gas temperature increase, while the CO ice abundance drops to zero and the CO enriched gas layers extend all the way to the midplane. The complex interplay between the disk dust structure, gas and dust temperature, and CO abundance underscores the valuable role of thermochemical modelling for interpreting line emission from disks.

\subsection{Parameter space}
\label{ssec:parameter_space}
We further explore the effects of varying various key parameters for the fiducial model to demonstrate how well constrained by the observations the fiducial model is and investigate the effect on the radiation field, gas and dust temperature, and CO abundance within the cavity.

\subsubsection{Effect of Inner Dust Ring}
\label{ssec:inner_disk}

\begin{figure*}[htb]
    \centering
    \includegraphics{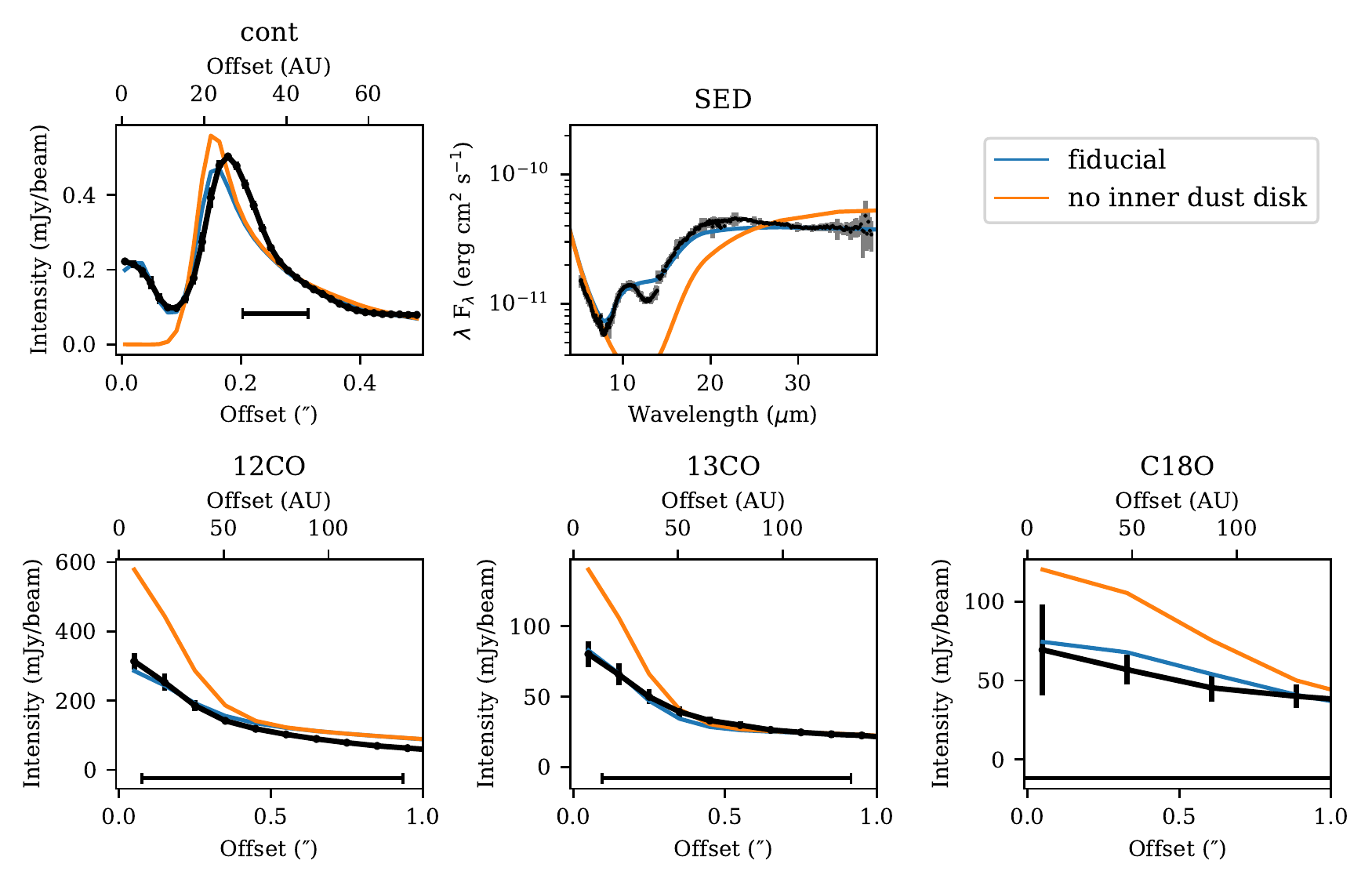}
    \caption{Comparison of the fiducial DM Tau DALI model and a model with no inner dust disk ($\delta_\text{dust1}=\delta_\text{dust2}=10^{-50}$ with the DM Tau observations, as described in Section \ref{ssec:inner_disk}. Top Row: deprojected and azimuthally averaged mm continuum intensity, and SED zoomed in on the near to mid infrared region. Middle Row: deprojected and azimuthally averaged intensity for the moment 0 maps of the $^{12}$CO, $^{13}$CO and C$^{18}$O lines.}
    \label{fig:inner_disk_comp_obs}
\end{figure*}

Many transition disks are observed to have an inner disk or ring within the central cavity \citep{Francis2020}, which is expected to effect the survival of CO in the cavity by shielding the gas in the cavity from stellar radiation \citep{Bruderer2013}. Here, we explore the effect of the inner dust ring at 3.5 au in DM Tau by removing it entirely from our fiducial model, i.e., setting both $\delta_\text{dust1}$ and $\delta_\text{dust2}$ to effectively zero values of 10$^{-50}$. The effect on the observations is shown in Figure \ref{fig:inner_disk_comp_obs}. The peak of the dust continuum emission at 21 au is $\sim 50\%$ brighter while the emission in the outer disk beyond $\sim 40$ au is unchanged; the SED is entirely missing emission from silicates at 10 $\mu$m and now peaks in the mid to far-infrared rather than at $\sim 20$ $\mu$m. The lack of an inner disk causes these effects by allowing more stellar radiation to reach the cavity edge at 21 au and removing the hot dust component in the cavity, as shown in Figure \ref{fig:inner_disk_comp_dali}. The CO emission is brighter in the cavity, particularly for the more optically thick $^{12}$CO and $^{13}$CO isotopologues, which reach 2-3 times higher temperatures at their $\tau=1$ surfaces; in the outer disk the CO emission is largely unchanged. Without an inner disk to shield the disk midplane from radiation, the CO freezeout zone is entirely absent in the cavity, and its onset begins a few au further out in the disk from the 21 au cavity edge than in the fiducial model. In contrast to the results of DALI modelling by \cite{Bruderer2013}, we find CO is not photodissociated within the dust cavity without an inner disk, which we discuss further in Section \ref{sec:disc}.

\subsubsection{Dust surface density}
\label{ssec:dust_surf}
The dust structure within the cavity of a transition disk can play an important role in regulating the radiation field throughout the disk and is typically constrained by the mm continuum emission and SED. We therefore consider the effect of varying parameters related to the dust structure within the cavity, including $r_\text{inner}$, $\delta_\text{dust1}$, and $\delta_\text{dust2}$. 

\begin{figure*}[htb]
    \centering
    \includegraphics{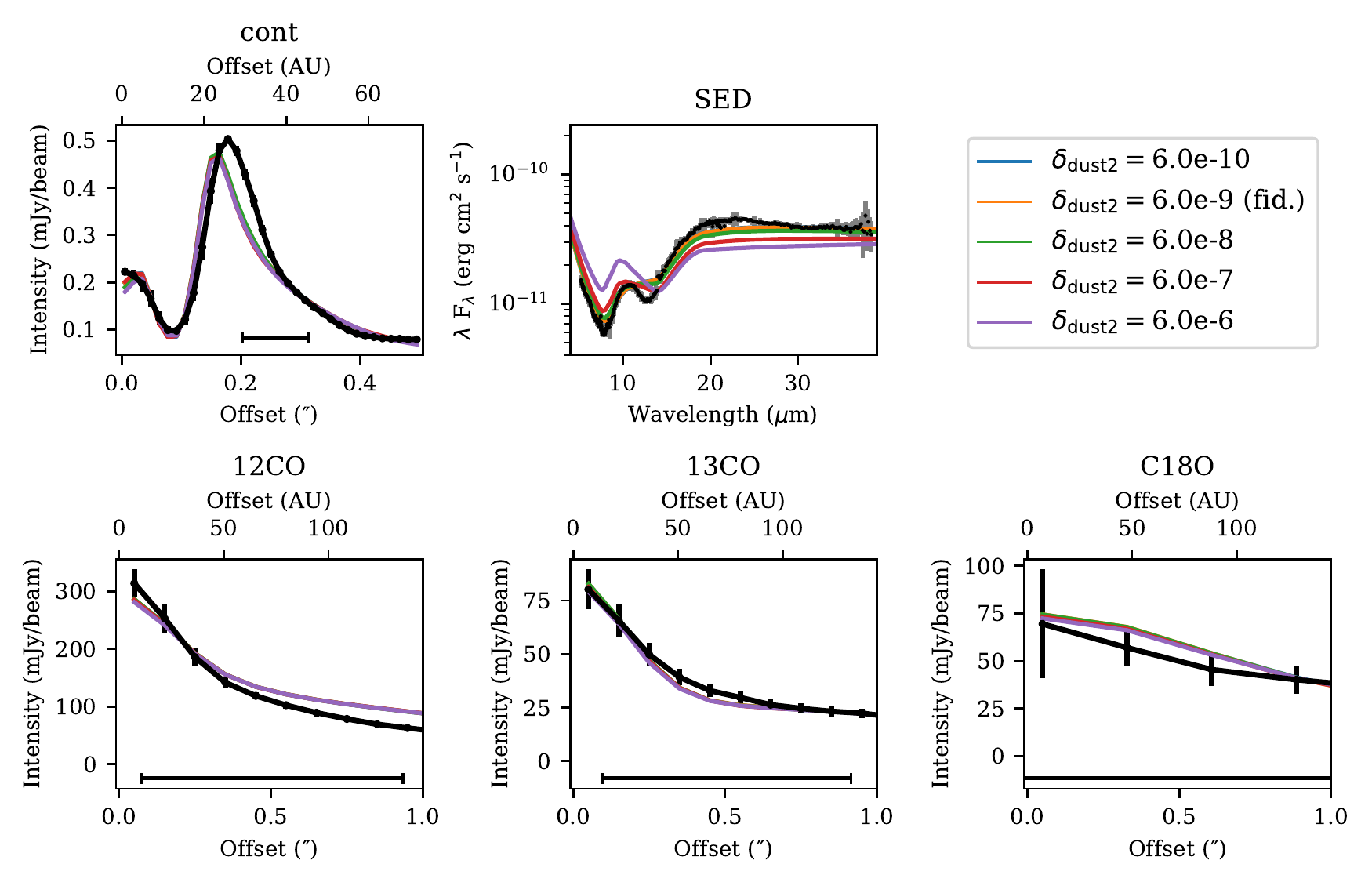}
    \caption{As Figure \ref{fig:inner_disk_comp_obs}, but for the effects of changing the dust depletion within 3.5 au in the DALI model described in Section \ref{ssec:dust_surf}.}
    \label{fig:ddust2_comp_obs}
\end{figure*}

In Figure \ref{fig:ddust2_comp_obs}, we show the effect on the observables of varying the dust surface density from the sublimation radius to 3.5 au by changing $\delta_\text{dust2}$. Only the SED is significantly affected, demonstrating its utility in constraining the dust distribution at distances from the host star that can not be resolved by ALMA. Reducing $\delta_\text{dust2}$ has no significant effect on the observables, indicating that the influence of the dust within 3.5 au in the fiducial model is negligible; the larger effect on the radiation field throughout the disk comes from the optically thick dust in the inner ring at $\geq$ 3.5 au. The near to mid-infrared emission is only visibly changed when $\delta_\text{dust2}$ is increased by two to three orders of magnitude above the fiducial, which produces too much emission at wavelengths of $\sim10$ $\mu$m, and too little in the mid-infrared. This is the result of creating a hot dust component within 3.5 au which emits in the near-infrared and shields the dust behind it, reducing the mid-infrared emission from cooler dust, as shown by the plots of the dust temperature and UV flux in the inner disk in Figure \ref{fig:ddust2_comp_dali}.

\begin{figure*}[htb]
    \centering
    \includegraphics{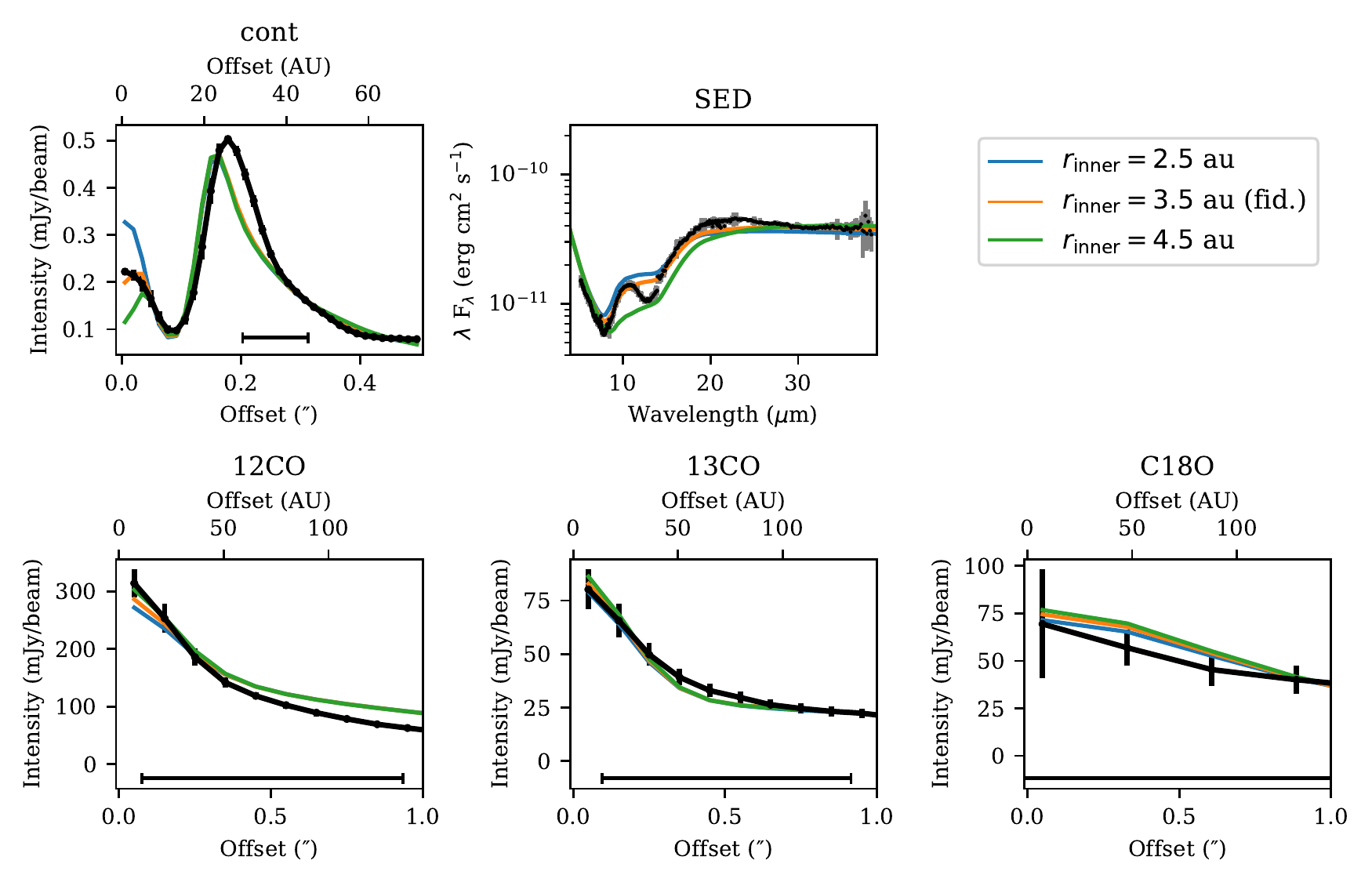}
    \caption{As Figure \ref{fig:inner_disk_comp_obs}, but for the effects of changing the radius of the inner edge of the interior dust ring ($r_\text{inner}$) in the DALI model described in Section \ref{ssec:dust_surf}.}
    \label{fig:rinner_comp_obs}
\end{figure*}

In Figure \ref{fig:rinner_comp_obs}, we show the effect of varying the radius of the inner edge of the interior dust ring ($r_\text{inner}$) while keeping the dust surface density scaling and depletion factors fixed. The ALMA observations are sufficiently high resolution for the mm continuum to be very sensitive to this radius, with values both 1 au larger and smaller than the fiducial 3.5 au ruled out. The observed near-infrared excess of the SED is slightly higher for the the 2.5 au case but underproduced in the 4.5 au case, suggesting that $r_\text{inner}$ is well constrained overall. A larger value of $r_\text{inner}$ causes an increased gas temperature at radii $<r_\text{inner}$ (Figure \ref{fig:rinner_comp_dali}), resulting in a slight change in the model CO emission within the 21 au cavity, however, this is poorly constrained by the CO observations due to the lack of resolution. 

\subsubsection{Gas surface density}
\label{ssec:gas_comp}

The structure of the gas surface density within the 21 au cavity of DM Tau can provide important input for models of planet disk interaction and accretion flow in the cavity. Here, we vary $\delta_\text{gas1}$ and $\delta_\text{gas2}$ within our fiducial DALI model to assess how well the radial CO emission profiles can constrain the gas surface density, and compare the results with the gas surface density implied by the accretion rate for an $\alpha$-disk model. 

\begin{figure*}[htb]
    \centering
    \includegraphics{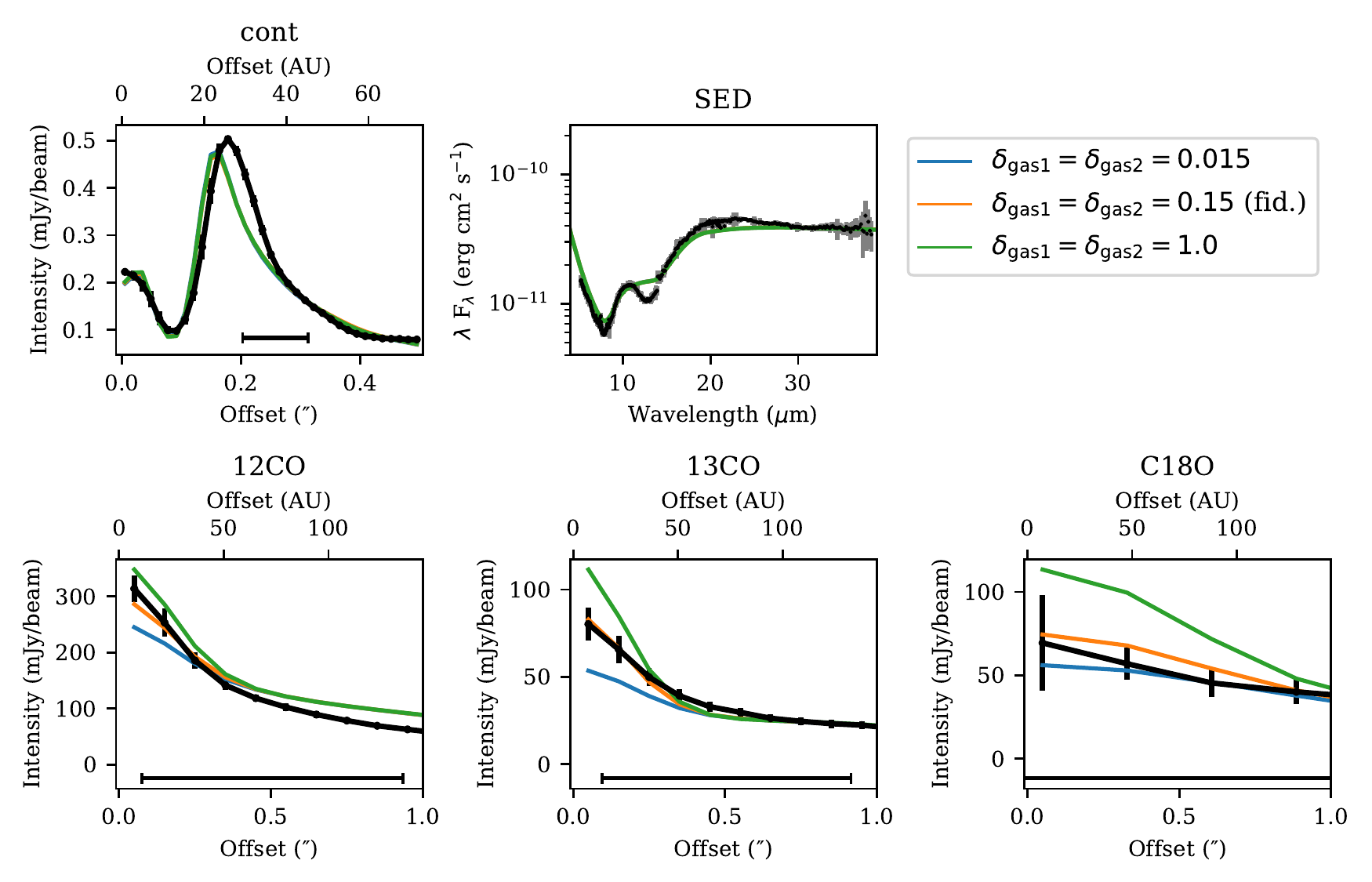}
    \caption{As Figure \ref{fig:inner_disk_comp_obs}, but for the effects of changing the gas surface density depletion within 21 au in the DALI model described in Section \ref{ssec:gas_comp}.}
    \label{fig:delta_gas_12_comp_obs}
\end{figure*}

We first vary $\delta_\text{gas1}$ and $\delta_\text{gas2}$ together to test the effect of a 10x deeper gas cavity than the fiducial model and a full disk with no gas gap. The comparison with the observables is shown in Figure \ref{fig:delta_gas_12_comp_obs}. There is no significant effect on the dust continuum emission or SED as the gas opacity is effectively zero with the PAHs removed from our model. 
Both the case of a full disk and a 10x depleted gap are ruled out by the CO emission, particularly by $^{13}$CO. In principle, the less abundant $^{13}$CO and C$^{18}$O isotopologues should be more optically thin than $^{12}$CO and therefore more sensitive to gas surface density changes in the 21 au cavity, however, the optical depth of each line is dependent on the gas density, temperature and CO abundance over the disk. In Figure \ref{fig:delta_gas_12_comp_dali} we show the gas temperature and CO abundance for the 21 au cavity and zoomed in on the inner disk with the $\tau=1$ surface of each CO line overlaid. $^{12}$CO is optically thick well above the midplane at all locations in the disk regardless of the gap depth. Nonetheless, $^{12}$CO emission is still sensitive to the gas surface density change in the gap because of the changing height of the $\tau=1$ surface and vertical temperature gradient in the disk. This is also the case for the $^{13}$CO emission for the fiducial and full disk models, which becomes optically thick above the midplane throughout the 21 au cavity. The C$^{18}$O emission is optically thin for the fiducial model and the 10x depleted gap, however, for a full disk it too becomes optically thick within the 21 au cavity.

\begin{figure*}[htb]
    \centering
    \includegraphics{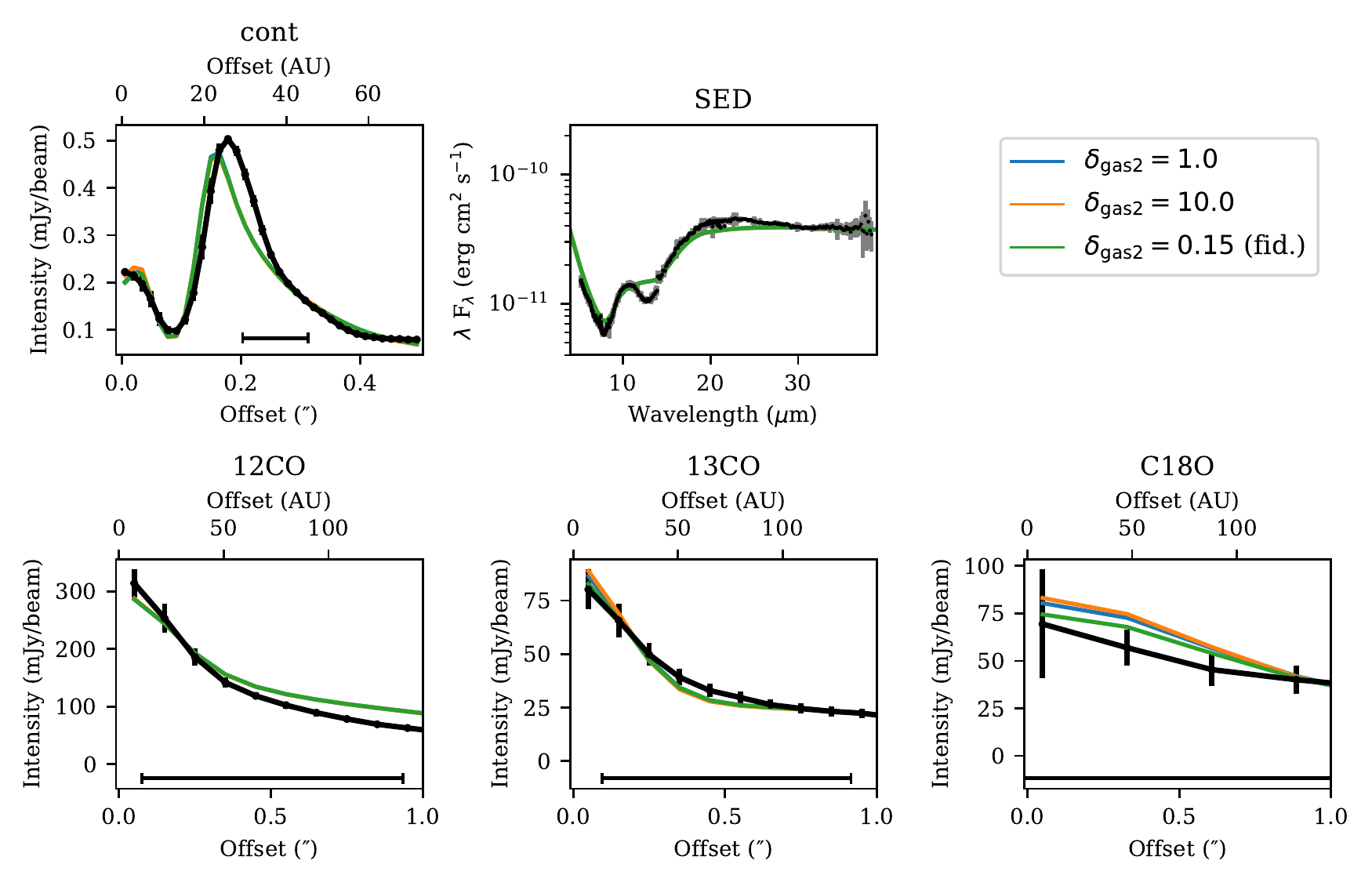}
    \caption{As Figure \ref{fig:inner_disk_comp_obs}, but for the effects of changing the gas surface density enhancement in the inner disk in the DALI model described in Section \ref{ssec:gas_comp}.}
    \label{fig:delta_gas_2_comp_obs}
\end{figure*}

We now increase $\delta_\text{gas2}$ by factors of 10 and 100 while fixing $\delta_\text{gas1}$ to the fiducial value to examine how sensitive the observations are to changes in the amount of gas in the innermost disk, which feeds accretion onto the host star. The results of this are shown in Figure \ref{fig:delta_gas_2_comp_obs}. Besides a small change in the model C$^{18}$O intensity --- which is not distinguishable with the present sensitivity and resolution of the observations --- no detectable change in the observables is seen. With deeper ALMA line observations resolving the inner disk, C$^{18}$O may be suitable as a tracer of the gas density, however, we note that its emission becomes optically thick above the disk midplane when there is an enhancement in the gas surface density (Figure \ref{fig:delta_gas_2_comp_dali}), indicating that careful modelling of the inner disk gas temperature would be required as well. 

The accretion rate onto DM Tau of $\sim 10^{-8.3}~\text{M}_\odot \text{yr}^{-1}$ \citep{Manara2014} can be translated into a gas surface density if we assume that the disk is well described by an $\alpha$-viscosity model. In this case, the surface density is related to the accretion rate by \citep{Manara2014}

\begin{equation}
\label{eqn:sigma_gas_mdot}
\Sigma_g(r)= \frac{\dot{M} 2 m_p}{3\pi\alpha k_BT(r)}\sqrt{\frac{GM_*}{r^3}}.
\end{equation}

\begin{figure}[htb]
    \centering
    \includegraphics{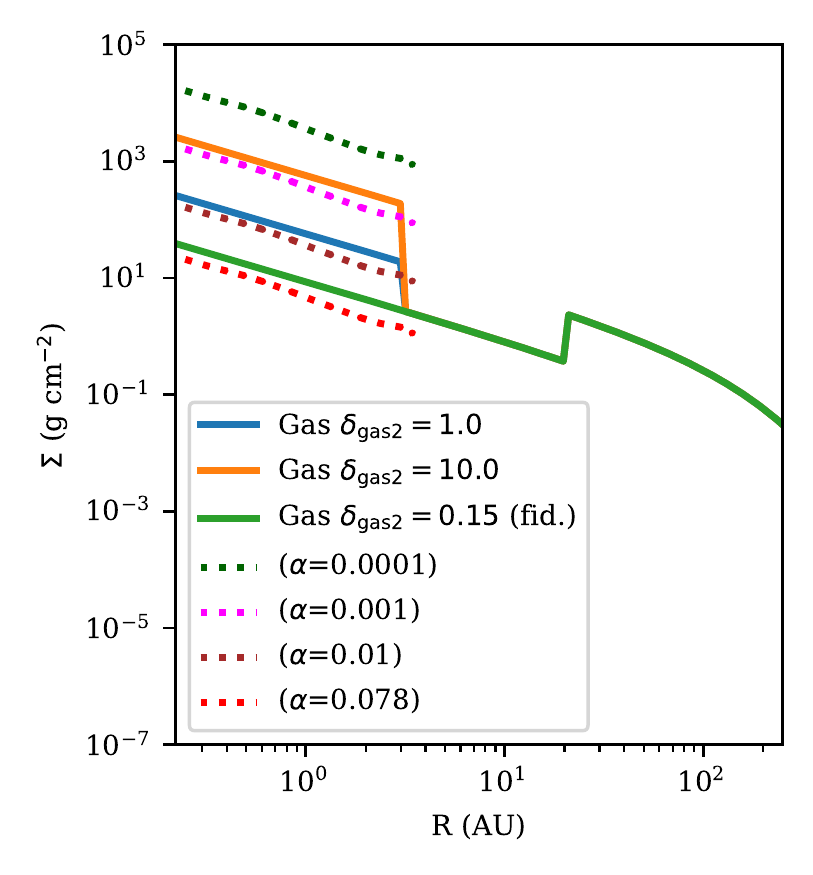}
    \caption{Comparison of the gas surface density for the input DALI models with varying enhancements in the inner disk with the gas surface density estimated from accretion rate with different alpha.}
    \label{fig:dgas_inner2_surfdens}
\end{figure}

This relationship implicitly assumes that the disk is vertically isothermal, whereas our DALI modelling provides a vertical temperature structure throughout the disk. To obtain a single representative temperature at each radius, we calculate the vertically averaged gas temperature weighted by the gas density for the fiducial model. The value of $\alpha$ is poorly constrained by observations for most disks, but is typically assumed to be low ($ \lesssim 10^{-3}$). DM Tau is one of few disks where turbulent broadening has been detected through high velocity resolution observations of the $^{12}$CO line \citep{Flaherty2020}, yielding a turbulent broadening of 0.25-0.33 c$_s$, or equivalently, $\alpha \sim (\delta v_\text{turb}/\text{c}_s)^2 = 0.078 \pm 0.02$. However, the $^{12}$CO emission arises from well above the midplane of DM Tau, and their observations do not resolve the inner disk, so the turbulence observed here may not be representative of the bulk of the accreting gas near the star. We thus plot the surface density of the gas derived from Eqn \ref{eqn:sigma_gas_mdot} and the fiducial model gas temperature assuming $\alpha=0.078,10^{-2},10^{-3},10^{-4}$, and compare it to the gas surface density structure for the fiducial model and those with an enhanced $\delta_\text{gas2}$ in Figure \ref{fig:dgas_inner2_surfdens}. The gas surface density in the inner disk for the fiducial model is consistent with $\alpha=0.078$ and the measured stellar accretion rate, however, a gas enriched inner disk is not ruled out by the CO observations, and is plausible if the $\alpha$ value of the accreting gas is $\lesssim 10^{-3}$. 

\subsubsection{Disk scale height}
\label{ssec:scale_height}
\begin{figure*}[htb]
    \centering
    \includegraphics{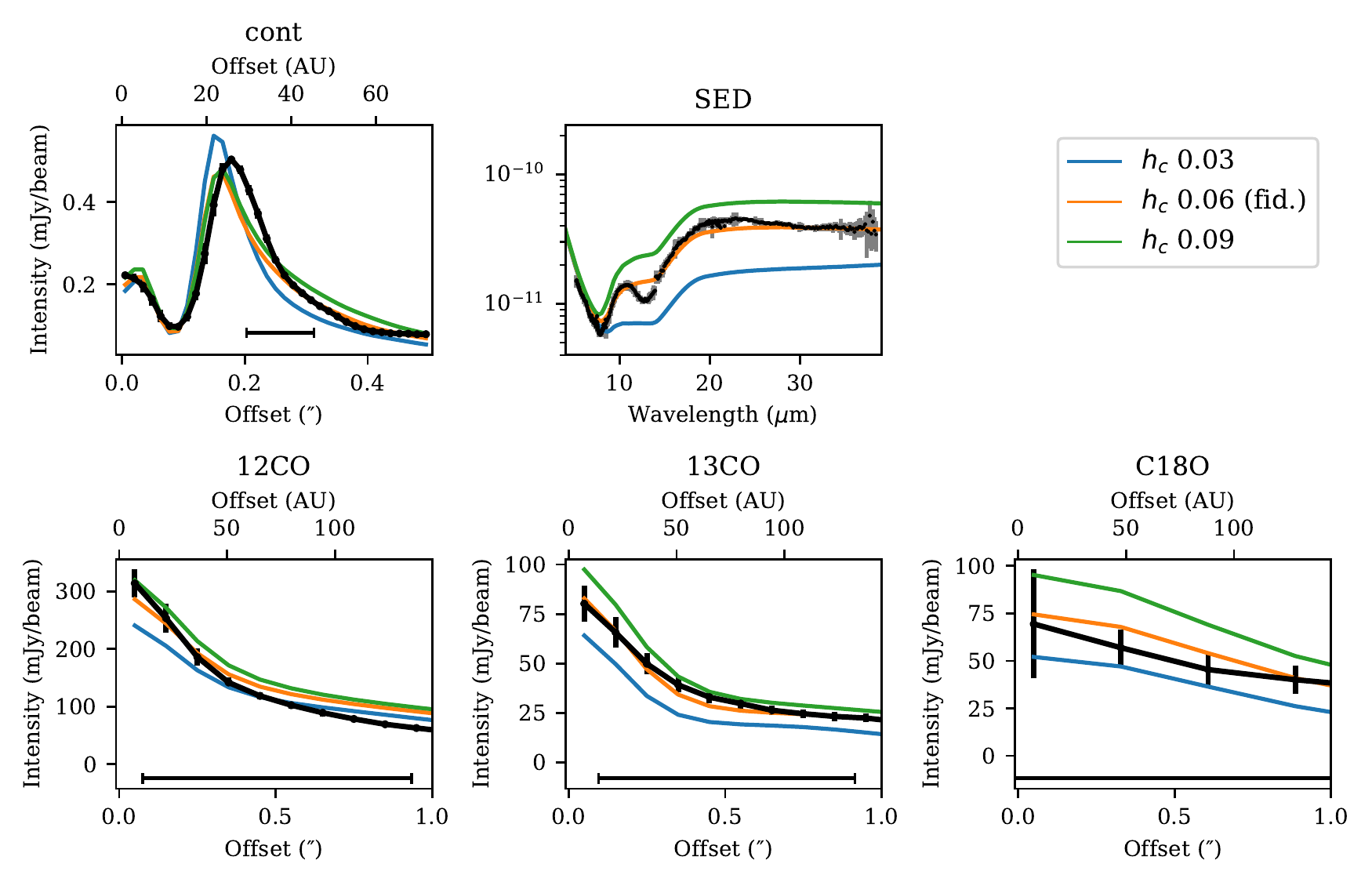}
    \caption{As Figure \ref{fig:inner_disk_comp_obs}, but for the effects of changing the disk scale height angle at the critical radius $h_c$ in the DALI model described in Section \ref{ssec:scale_height}.}
    \label{fig:h_c_comp_obs}
\end{figure*}

The scale height of the gas disk depends on the hydrostatic equilibrium balance between the gas pressure and gravity, and will affect the amount of stellar radiation absorbed by the disk surface as a function of radius. In our DALI model setup, the scale height of the small dust is the same as the gas, while that of the large dust grains is fixed to a fraction $\chi=0.2 h$ (Section \ref{ssec:vert_dust}). We thus test variations in the scale height of the disk by running DALI models with a flatter ($h_c$=0.03) and steeper ($h_c$=0.09) disk than the fiducial value ($h_c$=0.06), which affects both the dust and gas components of the disk. The effect on the model observations is shown in Figure \ref{fig:h_c_comp_obs}. With a larger scale height, the dust continuum is brighter in the outer disk and the peak of the emission at $\sim 21$ au is broader, the SED retains the same shape but is brighter at all near-infrared and longer wavelengths, and the emission for all CO isotoplogues is brighter throughout the disk, this applies conversely for a smaller scale height.
Corresponding plots of the DALI model gas temperature, CO abundance, and optical depth of the $\tau=1$ line are shown in \ref{fig:ddust2_comp_dali}. The location where the $\tau=1$ surface of the $^{13}$CO line reaches the freezeout temperature and drops to the disk midplane varies with the scale height, from radii of $\sim$ 25, 36, and 41 au for $h_c$ of 0.03, 0.06, and 0.09 respectively.

\subsubsection{Accretion rate}
\label{ssec:mdot}

\begin{figure*}[htb]
    \centering
    \includegraphics{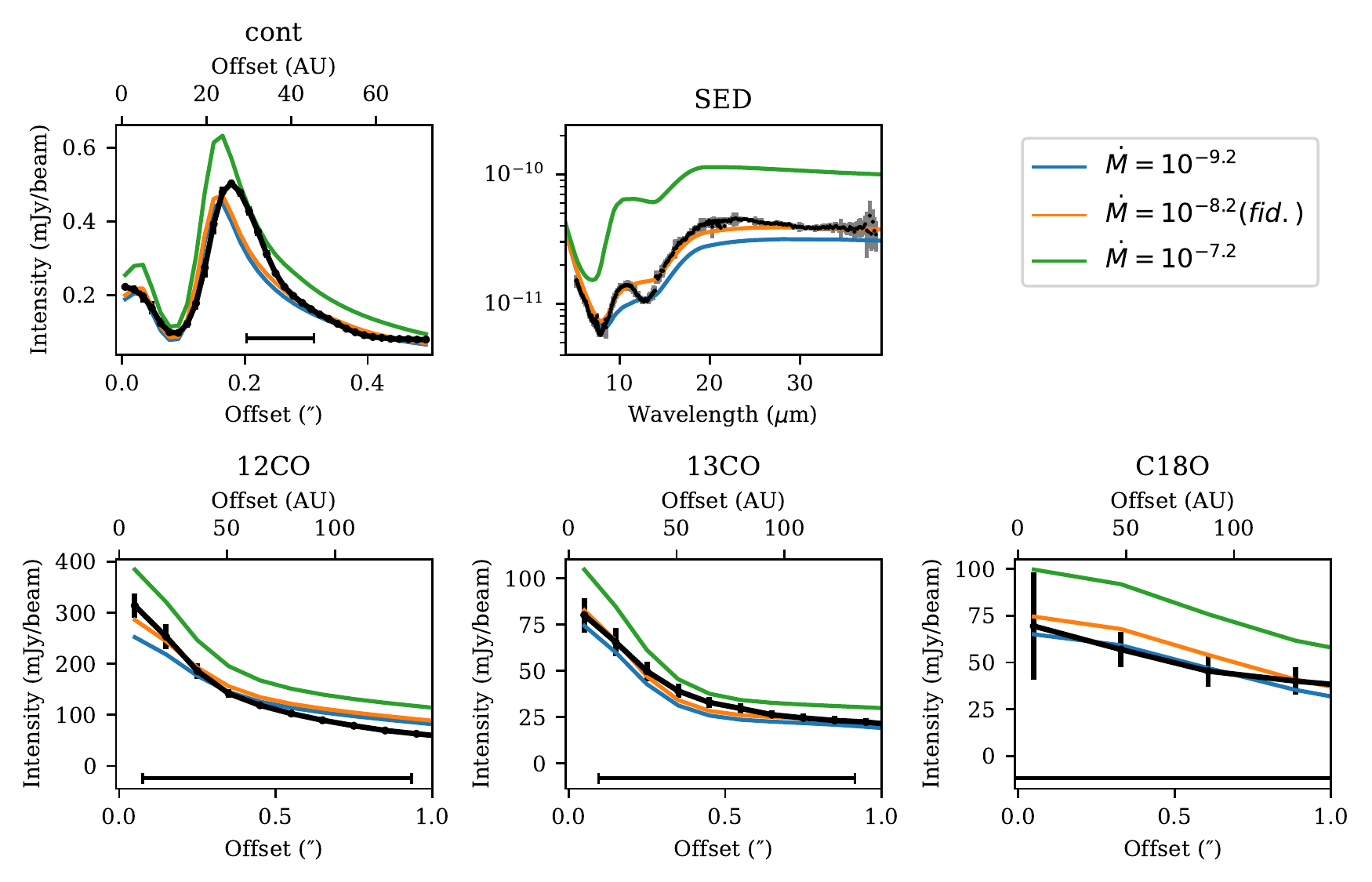}
    \caption{As Figure \ref{fig:inner_disk_comp_obs}, but for the effects of changing the stellar accretion rate in the DALI model described in Section \ref{ssec:mdot}.}
    \label{fig:mdot_disk_comp_obs}
\end{figure*}

While the accretion rate of DM Tau is reasonably well constrained by observations to within a factor of 3 \citep{Manara2014}, it is desirable to qualitatively examine how changes in the accretion rate affect the disk. In particular, 
the accretion luminosity is an additional source of radiation which can affect the thermal balance of the disk, and therefore the emission from the CO isotopologues which we use to estimate the gas surface density. Here, we thus vary the accretion rate by a factor of 10, producing a proportional change in the accretion luminosity (Eqn \ref{eqn:lacc}), with the majority of the change in the optical to UV part of the stellar model. The effect on the observations is shown in Figure \ref{fig:mdot_disk_comp_obs}. A 10 times larger accretion rate would brighten the dust continuum at all radii, with the the cavity edge at 21 au and the outer disk particularly affected. The SED would become brighter at all infrared and longer wavelengths, with a proportionally larger near-infrared excess, and the CO emission would be almost uniformly brighter throughout the disk. The opposite effect is seen for a 10 times lower accretion rate, however the difference with the fiducial is much smaller, particularly for the dust continuum. This effect is the result of contribution to the stellar model SED from the accretion luminosity in the optical-UV being a relatively small fraction of the emission from the star itself. In the DALI model itself (Figure \ref{fig:mdot_comp_dali}), with a higher accretion rate the UV flux is seen to penetrate deeper into the dusty layers of the disk, reducing the size of the midplane CO freezeout zone and increasing the size of the hot gas layer in the disk atmosphere. Overall, the observed changes in the CO emission with varying accretion rates highlight the importance of including accretion in the stellar model if accurate estimates of the gas structure are to be obtained.

\subsubsection{PAHs}
\label{ssec:pahs}

\begin{figure*}[htb]
    \centering
    \includegraphics{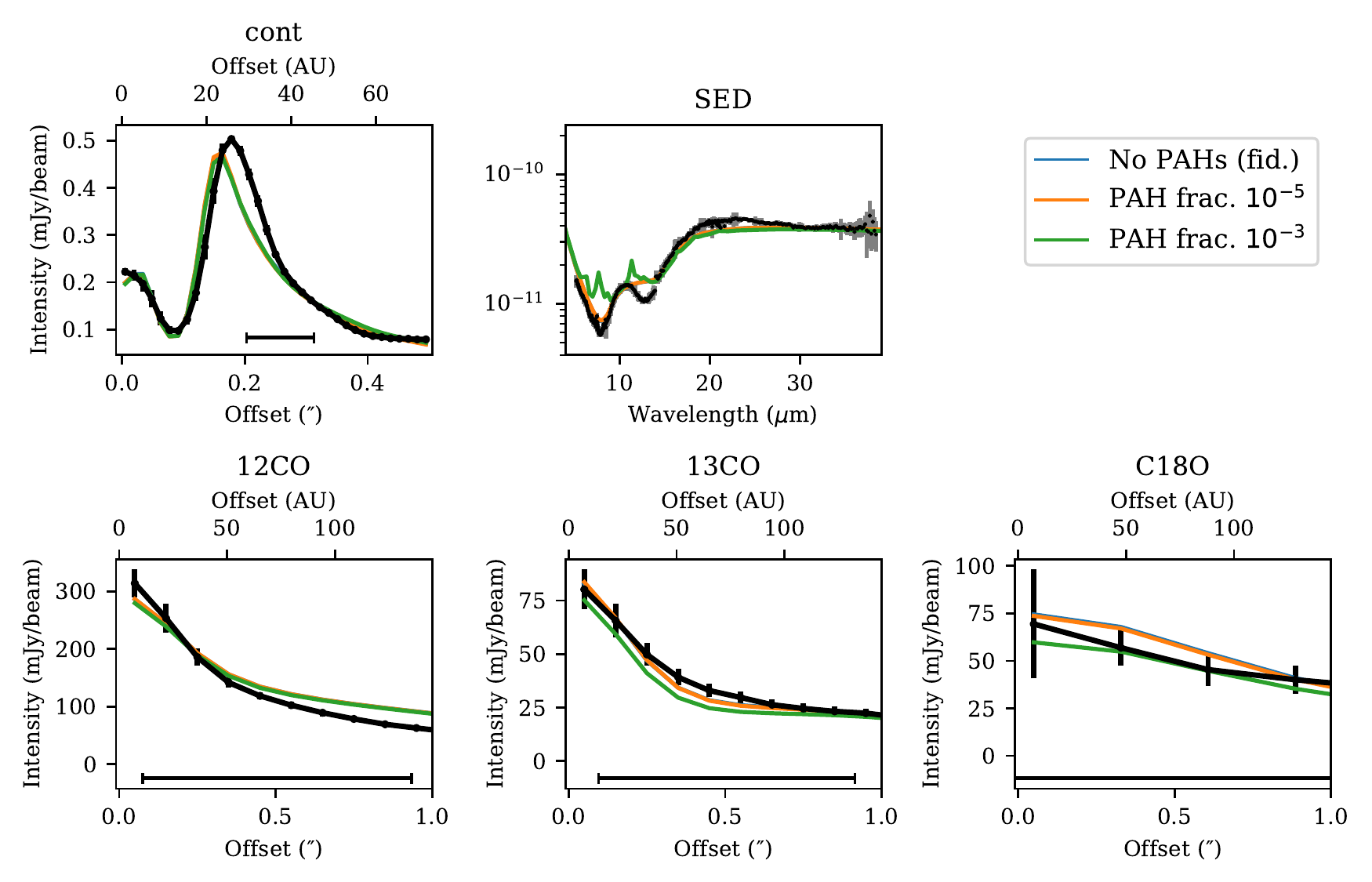}
    \caption{As Figure \ref{fig:inner_disk_comp_obs}, but for the effects of changing the global PAH abundance with respect to the ISM in the DALI model described in Section \ref{ssec:pahs}.}
    \label{fig:pah_comp_obs}
\end{figure*}

PAHs are a potential source of optical-UV opacity, gas heating, and H$_2$ formation \citep{Bruderer2013}, and would thus be expected to affect the gas temperature and CO abundance towards protoplanetary disks. PAH emission features are rarely detected (a lower limit of $\sim 8$ \%) towards T Tauri stars; in systems with a detection, a PAH abundance of 10-100x less than the ISM abundance is found, however, PAH emission was predicted to be undetectable in disks with host star  $T_\text{eff} < 4000K$ even for an ISM PAH abundance \citep{Geers2006}. In contrast, in our fitting of the fiducial DM Tau model, we found an extremely low PAH abundance fraction was required to avoid producing NIR PAH emission features, despite the relatively low $T_\text{eff}$ of DM Tau of 3580K. Here, we show the effect of increasing the $f_\text{PAH}$ with respect to the ISM from the fiducial $10^{-50}$ up to $10^{-5}$ and $10^{-3}$ on the observations in Figure \ref{fig:pah_comp_obs} and the inner disk of the DALI model in Figure  \ref{fig:pah_comp_dali}. The dust continuum is relatively unaffected by the changing PAH fraction, but for a $f_\text{PAH}$ of $10^{-3}$ or higher, NIR emission features which are not seen in the Spitzer IRS spectra appear in the model SED. With a higher PAH fraction, more of the UV emission is absorbed and more of the CO is frozen out towards the midplane. The more optically thin $^{13}$CO and C$^{18}$O emission is consequently fainter throughout the disk but not at a level we can distinguish within the sensitivity limits of the observations. The model PAH fraction is thus constrained to be $10^{-5}$ or lower, however we note that the opacity used for the PAHs in DALI assumes all PAHs are in the form of neutral $\text{C}_{100}\text{H}_{25}$; larger and/or charged PAH molecules may have a lower optical to UV opacity, thus allowing a higher PAH abundance, but a detailed exploration of the PAH composition is beyond the scope of this work.

The modelling of \cite{Geers2006} has suggested that protoplanetary disks around T Tauri stars could have ISM levels of PAHs without producing NIR emission features, however, their models did not include the additional UV flux from accretion, which is particularly significant for the relatively high accretion rate onto DM Tau. 

\section{Discussion}
\label{sec:disc}

\subsection{Summary of Modelling Results:}
The main findings from our DALI modelling of DM Tau are as follows:

\begin{itemize}
    \item Section \ref{ssec:inner_disk}: the inner disk of DM Tau absorbs a large amount of the radiation from the central star, resulting in a 2-3 times lower gas temperature at the $\tau=1$ of the optically thick CO lines surfaces within 21 au than would be seen without an inner disk. Furthermore, CO is able to survive in the cavity of DM Tau even without an inner disk providing shielding from photodissociation, in contrast to the DALI modelling results of \cite{Bruderer2013}.

    \item Section \ref{ssec:dust_surf}:  The mass of dust within 3.5 au quantified by the $\delta_\text{dust2}$ parameter is negligible, however, we do not perfectly reproduce the near-infrared SED shape, which is sensitive to the detailed structure of the inner ring.

    The value of $r_\text{inner}$ is well constrained to within an au of the fiducial value of 3.5 au by the combination of the mm continuum and SED fitting. The observed CO emission is not particularly sensitive to changes in either $\delta_\text{dust2}$ or $r_\text{inner}$.
    
    \item Section \ref{ssec:gas_comp}: the gas surface density within the 21 au dust cavity is depleted by a factor of $\sim 10$. Depletion by a factor of $\sim 100$ or an undepleted disk are both ruled out by our CO observations and modelling, however, a $\sim 10 \times$ depletion in combination with an enhancement in gas in the inner disk by a factor of $10-100$ could be consistent with the observed CO emission. A depletion factor of $\sim 10$ provides enough gas to sustain the observed accretion assuming an $\alpha$-disk model using the value measured from $^{12}$CO line turbulence of $0.078\pm 0.02$

    \item Section \ref{ssec:scale_height}: changes in the assumed scale height of the disk strongly affect the observed SED, and alter the dust continuum and CO emission in the outer disk by increasing or decreasing the amount of stellar radiation absorbed. The fiducial scale height angle of 0.06 is constrained by the SED within $\sim 0.01$.

    \item Section \ref{ssec:mdot}: an accretion rate 10x larger than the fiducial value produces a signficant increase brightness of the dust continuum, CO emission, and SED at all infrared wavelengths. An accretion rate 10x smaller than the fiducial only has a large observable effect through the overall decreased emission in the model SED.

    \item Section \ref{ssec:pahs}: DM Tau is extremely depleted in PAHs relative to the ISM value, the largest PAH fraction $f_\text{PAH}$ not ruled out by the lack of features in the near-infrared portion of the SED is $10^{-5}$ that of the ISM, assuming all PAHs are in the form of neutral $\text{C}_{100}\text{H}_{25}$ and that the distribution of PAHs through the disk is uniform.
\end{itemize}

\subsection{Modelling caveats and limitations of observations}
\label{ssec:caveats}

The gas temperature and chemical abundances solved for by DALI can have a high systematic uncertainty, as these results are tied to the details of the thermochemical modelling (e.g., the chemical reaction network and radiative transfer methods). A comparison of different thermochemical modelling codes suggests the gas temperature in the disk atmosphere may be uncertain up to a factor of 3 \citep{Rollig2007}, while typical low-J CO line fluxes are estimated to be accurate to about a factor of two \citep{Bruderer2013}.The chemical abundances in DALI are solved for steady state conditions, and do not include the effects of vertical or radial mixing by accretion or turbulence. Previous observations of $^{13}$CO in the outer disk have indicated a gas temperature below the freezeout temperature of CO, which has been argued as evidence of vertical mixing of gas from the warm disk midplane to the freezeout layer by turbulence \citep{Aikawa2006,Aikawa2007}. The assumed dust grain size distribution and the dust settling model used in DALI is parametric and not well constrained by observations. DALI does not include the effects of viscous heating from accretion, which may increase the gas and dust temperature in the inner disk. 

Self-shielding from photodissociation is less effective for the rarer CO isotopologues (i.e. $^{13}$CO, C$^{18}$O), resulting in their destruction deeper into the disk than the most abundant $^{12}$CO isotopologue and reduced line emission, though the magnitude of the effect depends on the stellar spectrum and dust structure (See \cite{Miotello2014}). An optional DALI mode to include the effect of isotope selective photodissociation is available, \citep{Miotello2014}, but we have not used it in our modelling due to the greatly increased computational cost. For the fiducial model, we have run one test case and found that the $^{13}$CO emission is unaffected but the C$^{18}$O line flux integrated over the disk is reduced by a factor of 3. While this implies that our gas surface density could be overestimated, this effect is largely important for colder regions of the outer disk where CO is frozen out. In our fiducial model, the gas temperature at the $\tau=1$ line does not drop below the range where CO freezes out until $\sim 45$ au for $^{13}$CO and $\sim 21$ au for C$^{18}$O (see Figure \ref{fig:dm_tau_fid_gastemptau1}). Therefore, our gas surface density estimates within the cavity should be unaffected by not including isotope selective photodissociation.

The CO to H$_2$ abundance ratio is a major source of uncertainty when estimating gas mass from CO observations, as CO can be destroyed by UV photons in the upper layers of the disk or frozen out onto dust grains in the midplane, reducing the column density. These processes are taken into account in our DALI modelling, however, gas masses measured from HD, which should have a constant abundance relative to H$_2$ and thus be an excellent tracer, are a factor of 5-100 times higher than the CO-derived gas masses \citep{McClure2016}. DM Tau happens to be one of only a few T Tauri disks with an HD detection from Herschel, which yields a mass range of $1.0 \times 10^{-2}$ to $4.7 \times 10^{-2}$ M$_\odot$. The integrated gas mass of our DALI model is $6.0 \times 10^{-3}$ M$_\odot$, which is lower by a factor of 1.6 to 7.8 compared to the HD measurement, suggesting there may indeed still be additional CO depletion processes at work in DM Tau, though this is difficult to constrain with the inherent uncertainties in the chemical network, and some of the disagreement may be reconciled if the outer disk gas mass is underestimated by a factor of a few as suggested by our model C$^{18}$O flux when isotope selective photodissociation is included.

Our biggest observational limitation is the lack of resolution in the cavity for the ALMA CO observations. This is due to the low signal on the longest baselines which results in significant tapering being required to obtain a detection of the more optically thin $^{13}$CO and C$^{18}$O; followup observations with a deeper integration time rather than a more extended ALMA configuration. The lack of resolution in our ALMA observations makes measurement of the gas surface density at scales of the inner disk of DM Tau poorly constrained, as shown in Section \ref{ssec:gas_comp}.

\subsection{Implications for dust gap opening and presence of planets}
\label{ssec:planets}

Whatever mechanism is responsible for the observed dust gaps of DM Tau must remain consistent with the shallow gas depletion and high accretion rate of $\sim 10^{-8.3}~\text{M}_\odot~ \text{yr}^{-1}$. Inside-out dispersal of transition disks has long been proposed as the origin of their dust gap, yet conventional photoevaporation models \citep[e.g.][]{Owen2010,Owen2011} are immediately ruled out for DM Tau, as they predict little to no stellar accretion. In contrast, the population synthesis models of \cite{Garate2021} find that gap opening by photoevaporation combined with a dead zone in the inner disk can produce a large fraction (63\%) of transition disks which are still accreting at rates $>  10^{-11}~\text{M}_\odot~ \text{yr}^{-1}$ after the gap is opened. The distribution of the accretion rates in their population synthesis models is sensitive to the dead zone radius and the decrease in $\alpha$-viscosity relative to the outer disk. In general, a greater fraction of disks with high accretion rates ($> 10^{-9}~\text{M}_\odot~ \text{yr}^{-1}$) is produced with a lower alpha viscosity in the dead zone and larger dead zone radius, as this results in a more extended inner disk which evolves on a longer viscous timescale and thus supplies accretion onto the star for a longer time. However, even in their most favourable case for producing disks with high accretion rates  with $\alpha = 10^{-4}$ and a dead zone radius of 5 au (Figure 5 of \cite{Garate2021}), very few disks ($\lesssim$ 0.1\%) are produced with accretion rates as high as DM Tau. DM Tau is thus either a true outlier, and/or possibly detected very recently after gap opening while the accretion rate is still high. While the low $\alpha=10^{-4}$ needed for a higher observed accretion rate in the dead zone models appears to be in disagreement with the $\alpha=0.078$ from turbulence measurements found by \cite{Flaherty2020}, their models for turbulent line broadening do not consider variations in $\alpha$ over the disk, nor does the $^{12}$CO line emission resolve the inner disk of DM Tau. The shallow gas depletion by a factor of $\sim 10$ in DM Tau is also difficult to reconcile with the \cite{Garate2021} models, which predict an extremely deep depletion ($> 10^{-3}$) only few Myr after gap opening. Furthermore, while the simulations of \cite{Garate2021} which include dust evolution and raytracing can produce a bright outer dust ring and an inner disk which is a factor of a few times fainter at mm wavelengths, the overall mm flux in their models is typically around 10 mJy, whereas the integrated flux in our ALMA band 6 images of DM Tau is much higher, $\sim 142$ mJy. \cite{Garate2021} note that for large and mm-bright transition disks, planets may thus still be required to produce pressure bumps and trap mm-sized dust grains even if photoevaporation is simultaneously opening a gap. Overall, the observed properties of DM Tau are thus difficult to explain using the photoevaporative models.

Large dust rings in transition disks have also been hypothesized to be the result of clearing or trapping in a pressure bump by giant planets. While observations of DM Tau intended to find planets in the gap (see Section \ref{sec:dm_tau}) have not provided a clear detection, translating observed fluxes or upper limits to a companion mass is difficult owing to our poor understanding of the properties of young giant exoplanets. In particular, the predicted planet mass for a given brightness varies considerably depending on whether the gravitational potential energy lost during planet formation is radiated away by accretion  \citep[``cold-start'' models; e.g.][]{Marley2007,Fortney2008} or partially retained as heat \citep[``hot-start''; e.g.][]{Baraffe2003}. Possible episodic accretion bursts \citep{Lubow2012,Brittain2020} and the likely presence of infrared-bright circumplanetary disks \citep{Zhu2015} further complicate the interpretation. Assuming a hot-start model of planet formation, the near-infrared Subaru observations of \cite{Uyama2017} place a typical upper limit on planet mass of $\sim 10$ M$_\text{J}$ at 70 au, however, their contrast limits are unlikely to be able to detect planets within the 21 au dust ring of DM Tau at $\le 0.15$\arcsec\ due to the significantly poorer contrast at short separations. The tentative detection of a companion at $\sim 6$ au using sparse aperture masking observations of \cite{Willson2016} was interpreted using the accreting circumplanetary disk models of \cite{Zhu2015} as planet and disk with $M_p \dot{M} = 10^{-5}~M_{J}^2 \text{yr}^{-1}$, which for an average circumplanetary disk accretion rate sufficient to form a 1 $M_J$ planet in 1 Myr of $10^{-9} M_\odot \text{yr}^{-1}$, is equivalent to a planet mass of 10 $M_J$. However, the detected emission may also be from the bright inner disk of DM Tau, as sparse aperture masking can result in extended emission being confused for point sources \citep{Currie2019,Blakely2022}. Overall, the presence of giant planets within the 21 au dust gap of DM Tau is not ruled out at the sensitivity limits of observations to date.

One of the hallmarks of gap opening by a giant planet is its effect on the gas gap structure. The width of the gas gap is predicted to be smaller than the dust gap \citep{Pinilla2012a}, which has been observed in other transition disk gaps by \cite{vanderMarel2016b}. However, the effective resolution of our ALMA observations is too low to identify this; the beam size of the $^{12}$CO and $^{13}$CO observations is $\sim 42$ au, and the dust cavity with radius 21 au is thus poorly resolved. The depth of the gas gap is sensitive to the planet mass, as shown by 2D hydrodynamical simulations of $\alpha$-disks \citep{Fung2014}. An empirical formula derived from these simulations relates the gas depth to the planet-star mass ratio $q$, the viscosity parameter $\alpha$, and disk scale height $h$:
\begin{equation}
\label{eqn:delta_gas_fung}
\delta_\text{gas} = 4.7\times10^{-3}\left(\frac{q}{5\times10^{-3}}\right)^{-1}\left(\frac{\alpha}{10^{-2}}\right)^{1.26}\left(\frac{h}{0.05}\right)^{6.12}
.\end{equation}
This relation is only valid for $q=10^{-4}-10^{-2}$, higher planet-star mass ratios result in a gravitationally unstable disk in the simulations of \cite{Fung2014}. Using the $\alpha=0.078$ value derived from the turbulence measurements of \cite{Flaherty2020} and the fiducial model gas gap depth and scale height at 21 au of 0.15 and $h=h_c(r/r_c)^\psi = 0.042$ respectively yields $q=7.2 \times 10^{-4}$, which for the stellar mass of DM Tau of 0.39 M$_\odot$ is equivalent to $M_p = 0.29\, \text{M}_J$. However, this planet mass is highly uncertain, as equation \ref{eqn:delta_gas_fung} is extremely sensitive to the disk scale height, and was derived assuming a vertically isothermal disk. If we instead assume $h$ values at 21 au of 0.035 ($h_c=0.05$) or 0.049 ($h_c=0.07$) which are reasonably consistent with the fiducial SED, we arrive at planet masses of $M_p = 0.097 \text{M}_J$ and $M_p = 0.76 \text{M}_J$ respectively.

The sub-Jovian planet mass estimated above is at odds with the dust structure of DM Tau and predictions from hydrodynamical simulations of dust evolution in a disk with a giant planet in the gap. \cite{Pinilla2012a} find that a $1 \text{M}_J$ planet with $\alpha=10^{-2}$ opens a shallow gas gap with only a depletion in gas surface density of a factor of $\sim 10$ (their figure 1, top left panel) relative to the outer disk, however, the pressure gradient is not strong enough to prevent the rapid radial drift of mm-size and smaller dust grains, resulting in a removal of most of the disk dust within a few Myr. Effective trapping of dust grains is possible with a $1 \text{M}_J$ planet if $\alpha$ is $<= 10^{-3}$, for a much more massive 10 $\text{M}_J$ planet the efficiency of trapping is not as dependent on alpha and a dust ring is formed even for $\alpha=10^{-2}$; but in either case the gas gap is expected to be depleted by $>10^2$. A narrower but deeper gas gap could be consistent with the ALMA CO observations, which poorly resolve the 21 au dust cavity and would also result in a proportionally higher planet mass from equation \ref{eqn:delta_gas_fung}.

The bright inner dust ring of DM Tau could indicate that the dust trapping by a planet producing the 21 au outer dust ring is not completely efficient. Further hydrodynamical simulations with dust evolution and a giant planet have shown that for an $\alpha=10^{-3}$ disk and planet mass $> 5 \text{M}_J$, dust trapping is extremely efficient and an inner disk producing a near-infrared excess can only be maintained for 1 Myr \citep{Pinilla2016}. This implies that some combination of a lower planet mass, stronger disk turbulence, or a recently opened gap are needed to have an observable inner disk, however, exploring this would require detailed simulations that are beyond the scope of this work. 

\cite{Francis2020} quantified the properties of inner dust disks in transition disk gaps and found that the inner dust disk of PDS 70 - the only transition disk with confirmed planets in the gap - was an outlier for both its large size and much lower gas to dust ratio than other transition disks. This was hypothesized to be the combined result of planets only recently opening the gap and the inward radial drift of mm sized grains: with a newly opened gap, the supply of drifting mm grains to the inner disk will have only recently been cut off, and the dust grains in the inner disk will not have had sufficient time to drift inwards, resulting in a large and relatively dust rich inner disk. Furthermore, the young planets in the gap should be brighter and more easily detected, explaining why planets have not been identified towards other transition disks despite similar deep searches (\citep[e.g.][]{Maire2017,Zurlo2020}. A characteristic outer radius of the inner ring of DM Tau was measured by \cite{Francis2020} using a Gaussian fit to be $7.5 \pm 0.3$ au. This radius is relatively large within the sample, but the gas to dust ratio inferred from the accretion rate and assuming an $\alpha$-disk model with $\alpha=10^{-3}$ is $\sim 10^{4}$, which was similar to most other inner disks. Our fiducial DALI model has a gas to dust ratio in the inner disk of $\sim 50$, which would suggest it is a possible outlier in the same sense as PDS 70, however, the gas content of the inner disk derived from the CO observations is uncertain by several orders of magnitude (see Section \ref{ssec:gas_comp}). ALMA observations resolving the inner disk are required to better constrain it's gas content and deduce if it is the result of recent gap opening.

\subsection{The inner disk of DM Tau}
As we showed in Section \ref{ssec:inner_disk}, the inner disk plays an important role in reducing the radiation field and temperature within the 21 au cavity, and thus the brightness of the optically thick CO lines. The seminal DALI modelling of transition disks by \cite{Bruderer2013} found that for a gas depleted transition disk cavity ($\delta_\text{gas} \leq 10^{-4}$), an inner dust disk and PAHs within a transition disk cavity were essential to ensuring the survival of molecular H$_2$ and CO, as the dust and PAHs provide both shielding from photodissociation and a site for efficient formation of H$_2$ and other molecules. On the other hand, we find that CO is abundant within the 21 au cavity even without any inner disk. This apparent discrepancy is likely the result of the significantly lower levels of gas depletion required in our DM Tau model. \cite{Bruderer2013} note that with higher gas densities, H$_2$ can alternatively form by the reaction H$^{-1}$ + H $\rightarrow$ H$_2$ + e$^{-}$, and provide a pathway to CO formation. This highlights the importance of considering the dust content of the inner disk and overall gas structure in accurately translating observations of CO to a gas surface density.

\begin{figure}
    \centering
    \includegraphics{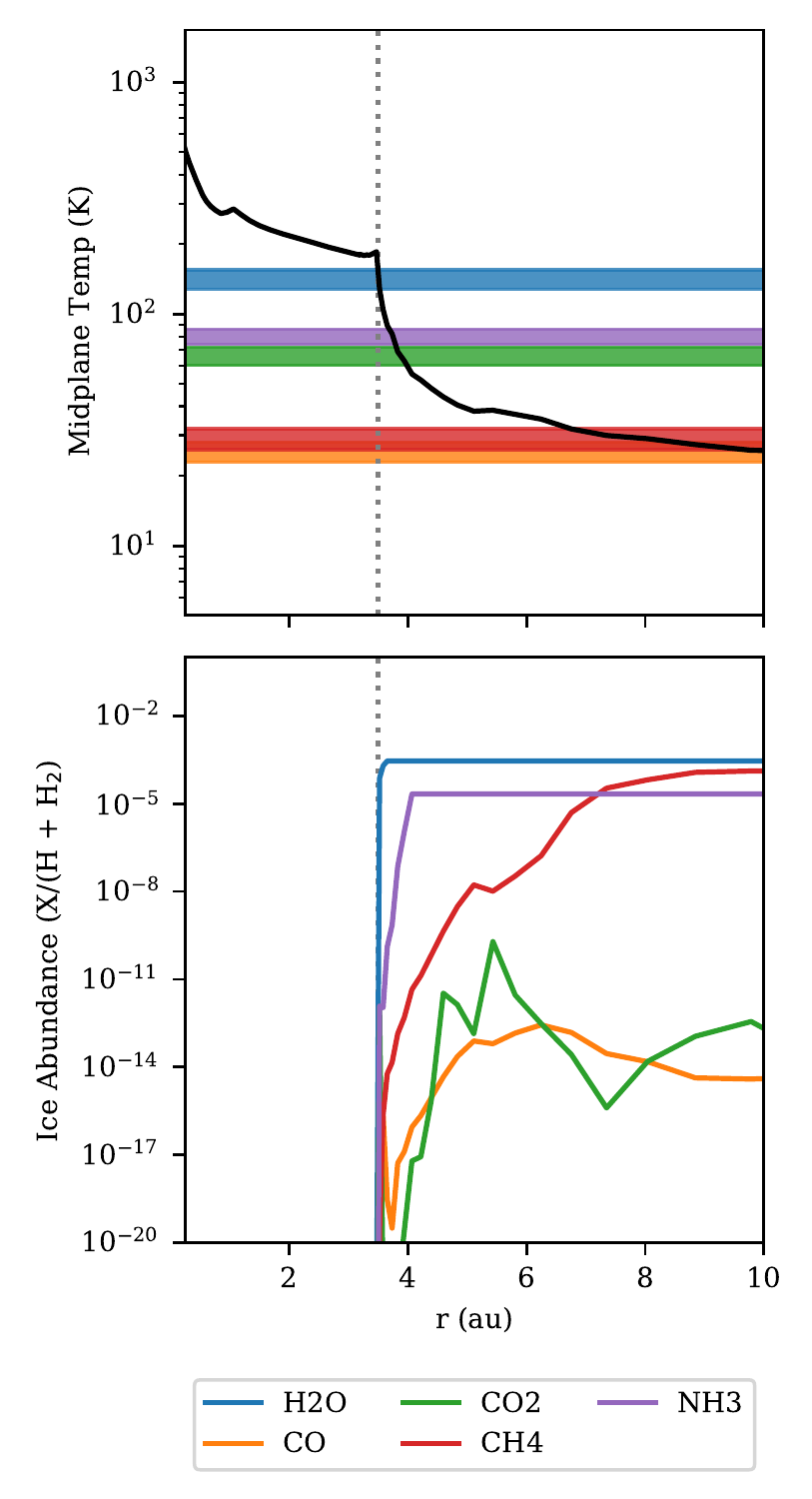}
    \caption{Midplane ($z=0$) temperature (top panel) and abundance of various volatile species locked into ices (bottom panel) in the fiducial DM Tau model within 10 au. X/H+H$_2$ denotes the ratio of the number densities of ice X and the total gas density. N.B.: the gas and dust temperature at the midplane are essentially identical in this region of the disk.}
    \label{fig:fiducial_snowlines}
\end{figure}

In our modelling of the SED and dust continuum, the radius for the inner dust cavity ($r_\text{inner} \sim 3.5$ au) in our fiducial model agrees with previous radiative transfer modelling efforts by \cite{Calvet2005} ($\sim 3$ au) and \cite{Hashimoto2021} ($3.22^{+0.31}_{-0.12}$). The origin of the inner dust cavity is unclear and difficult to constrain without higher resolution observations of the inner disk. Clearing of disk material and dust trapping by one or more planets within 3.5 au could plausibly explain an inner dust ring, but raises the question of how this would affect the accretion onto the host star. The collisional properties of dust grains such as the likelihood of fragmentation, sticking, and bouncing, may change significantly depending on the amount and composition of ices on their mantles \citep{Bauer2008,Birnstiel2010,Testi2014}. The presence of snowlines at or near the 3.5 au radius could thus lead to the removal of mm-sized grains through destruction or aglomeration into larger bodies, producing an apparent inner ring. Modelling of these processes in DM Tau is beyond the scope of this paper, however, our DALI model allows us to explore the locations where common ices may sublimate in the inner disk. In Figure \ref{fig:fiducial_snowlines}, the midplane temperature of the dust and gas, sublimation temperature ranges of \cite{Zhang2015}, and abundance relative to Hydrogen of various ices at the midplane is shown for the fiducial DM Tau model. Within most of the inner disk outside 3.5 au, the midplane is cold ($\sim 30$ K) and most ices should be frozen out. As the inner cavity edge at 3.5 au is approached, the midplane temperature rises to $\sim 200$ K, sufficient to remove essentially all of the ices on the grains. Any planets formed within 3.5 au would thus be expected to incorporate solids and gas which are respectively relatively depleted or enriched in volatiles.

\section{Conclusions}
\label{sec:conc}
In this paper, we have used thermochemical modelling with the DALI code in combination with ALMA images of the disk continuum emission, $^{12}$CO, $^{13}$CO, and C$^{18}$O lines, SED, and stellar accretion rate to constrain the properties of the gas cavity and inner disk of DM Tau. Our major conclusions are the following:

\begin{itemize}

    \item The gas surface density within the 21 au dust cavity is appears to show only a shallow depletion by a factor of $\sim 10$, which is consistent with estimates of the gas surface density in the inner disk from the stellar accretion rate and viscous $\alpha$-disk modelling, however, the gas density of the inner disk itself is poorly constrained and could be enhanced by a factor of $10-100$ relative to a shallow gap and still be consistent with our interpretation of the CO observations and the accretion rate.

    \item The dead zone + photoevaporation models of \cite{Garate2021} can produce transition disks with an inner dust disk, outer ring, and high accretion rate, but are still unlikely as an explanation for the gap in the DM Tau transition disk, as they do not reproduce the high stellar accretion rate or bright mm emission in the dust disk. However, non-viscous angular momentum transport in the inner disk and uncertainties in the model dust opacities may contribute to the disagreement between their models and the observations.
    
    \item The planet mass of $\lesssim 1 M_J$ implied by the shallow gas gap and $\alpha=0.078 \pm 0.02$ from turbulence measurements \citep{Flaherty2020}, is inconsistent with the larger planet masses required for dust trapping in a highly viscous disk. A deeper but narrower gap unresolved by our ALMA observations could allow for a higher planet mass consistent with dust trapping by a giant planet.

    \item The inner disk of DM Tau absorbs much of the stellar radiation, lowering the gas temperature in the cavity. In contrast to the DALI transition disk modelling of \cite{Bruderer2013}, CO is able to survive in the shallow gas cavity of DM Tau even without an inner dust disk providing shielding from photodissociation, which is likely the result of H$_2$ formation at higher gas densities by the H$^{-1}$ + H $\rightarrow$ H$_2$ + e$^{-}$ reaction, which provides an alternative pathway to CO formation.

    \item Our modelling of the inner disk indicates a $\sim 3.5$ au cavity, consistent with previous studies by \cite{Hashimoto2021} and \cite{Calvet2005}
    Most volatiles species are likely in the gas phase within 3.5 au cavity, and may thus be incorporated into planets forming within. 
    
    \item Our SED modelling indicates DM Tau is extremely depleted in PAHs relative to the ISM value, and has a PAH fraction of at most $10^{-5}$ that of the ISM, assuming that all PAHs are in the form of neutral $\text{C}_{100}\text{H}_{25}$ and that the PAH distribution is uniform throughout the disk. Unlike previous studies of PAH survival in transition disks \citep{Geers2006,Geers2007}, our modelling includes UV flux from accretion in assessing the PAH abundance, resulting in a lower inferred value. 
    
\end{itemize}

\begin{acknowledgments}
We thank the anonymous referee for their helpful comments on initial drafts of this paper.
This paper makes use of the following ALMA data: ADS/JAO.ALMA\#2013.1.00498.S, ADS/JAO.ALMA\#2017.1.01460.S. ALMA is a partnership of ESO (representing its member states), NSF (USA) and NINS (Japan), together with NRC (Canada), MOST and ASIAA (Taiwan), and KASI (Republic of Korea), in cooperation with the Republic of Chile. The Joint ALMA Observatory is operated by ESO, AUI/NRAO and NAOJ.


D.J.\ is supported by the National Research Council of Canada and by an NSERC Discovery Grant.
E.A.\ is supported by MEXT/JSPS KAKENHI grant No. 17K05399. H.B.L.\ is supported by the Ministry of Science and
Technology (MoST) of Taiwan (Grant Nos.\ 108-2112-M-001-002-MY3 and 110-2112-M-001-069-). Y.Y.\ was supported by NAOJ ALMA Scientific Research Grant Numbers 2019-12A.

\end{acknowledgments}

%

\vspace{5mm}
\facilities{ALMA, VLT (X-SHOOTER), Spitzer, GAIA}


\software{\texttt{astropy} \citep{AstropyCollaboration2013,AstropyCollaboration2018}, \texttt{CASA} \citep{McMullin2007}, \texttt{bettermomments} \citep{Teague2018}.}



\appendix
\section{Fiducial DALI Model Images and Residuals}
\begin{figure*}[htb]
    \centering
    \includegraphics{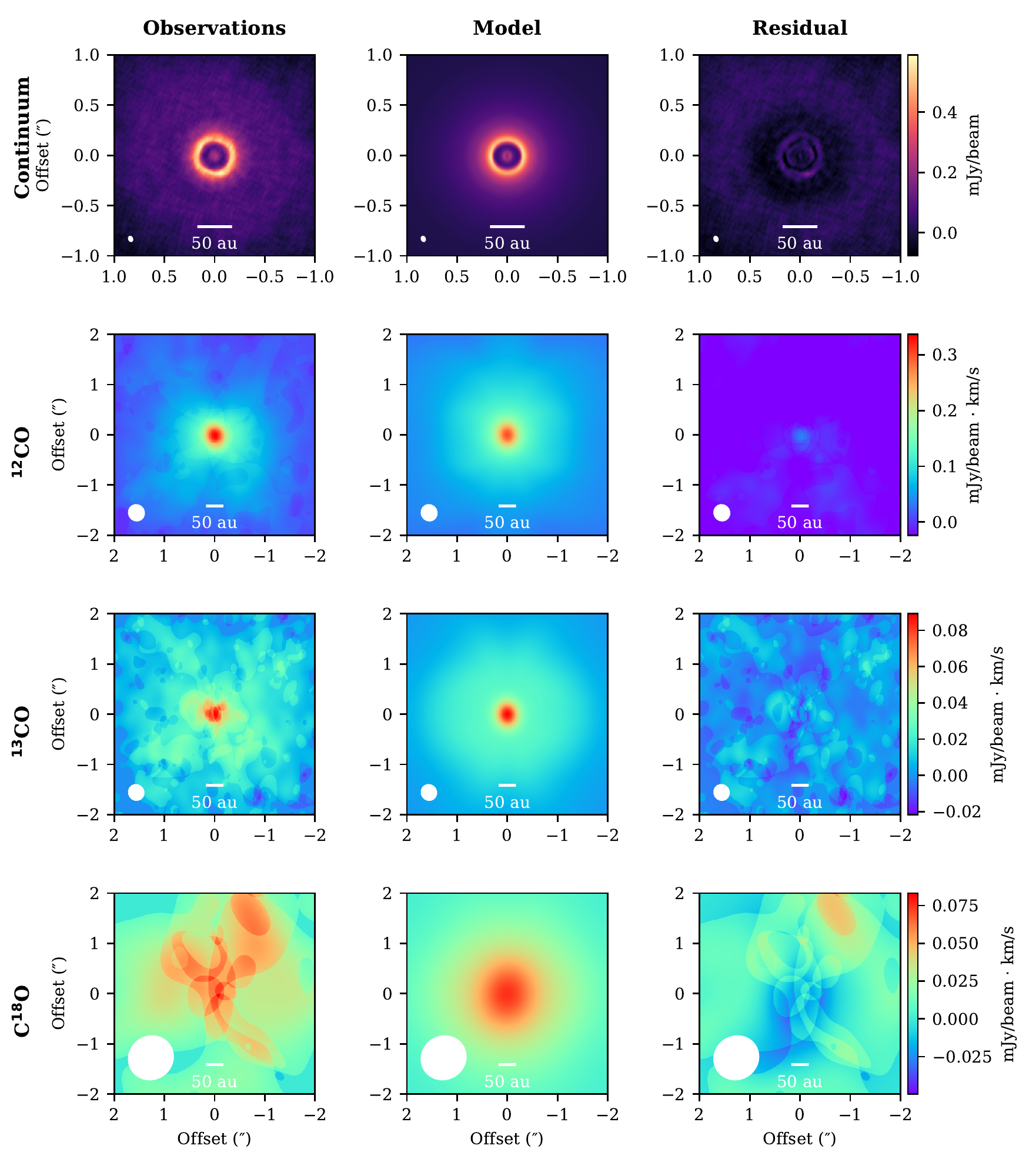}
    \caption{Observed (left column), fiducial DALI model (center column), and residual (right column) deprojected images for the 1.3mm dust continuum (first row) and $^{12}$CO, $^{13}$CO, and C$^{18}$O moment 0 maps (second to fourth rows). The beam for each image is shown in the bottom-left of each panel as a white ellipse.}
    \label{fig:dm_tau_fid_im_res}
\end{figure*}

In Figure \ref{fig:dm_tau_fid_im_res}, we show the deprojected ALMA dust continuum and moment 0 maps of the $^{12}$CO, $^{13}$CO, and C$^{18}$O J=2--1 lines, the corresponding fiducial DALI model images, and the residual images after subtracting the model from the observations. 

\section{DALI MODEL Full Disk}
\begin{figure*}[htb]
    \centering
    \includegraphics{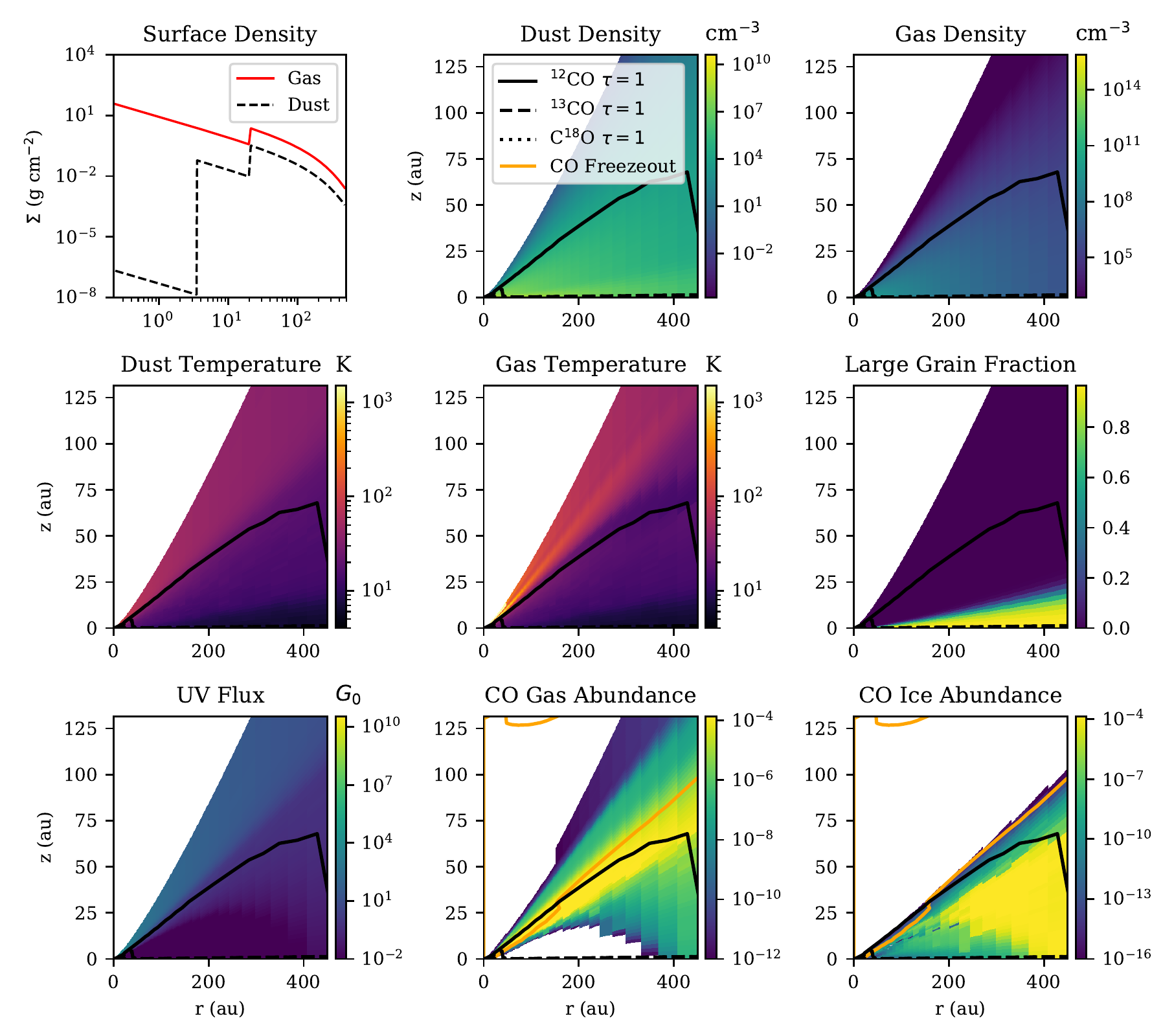}
    \caption{The same fiducial DALI model as Figure \ref{fig:dm_tau_fid_surf_dens_dali}, but showing the full model grid from from 0 to 500 au.}
    \label{fig:dm_tau_fid_dali_full_disk}
\end{figure*}

In Figure \ref{fig:dm_tau_fid_dali_full_disk}, the fiducial DALI model of DM Tau is shown over its full extent from 0 to 500 au for relevant DALI physical quantities (gas and dust density and tem-
perature, CO abundances, etc.)

\section{DALI Models Parameter Space Exploration}

\begin{figure*}[htb]
    \centering
    \includegraphics[]{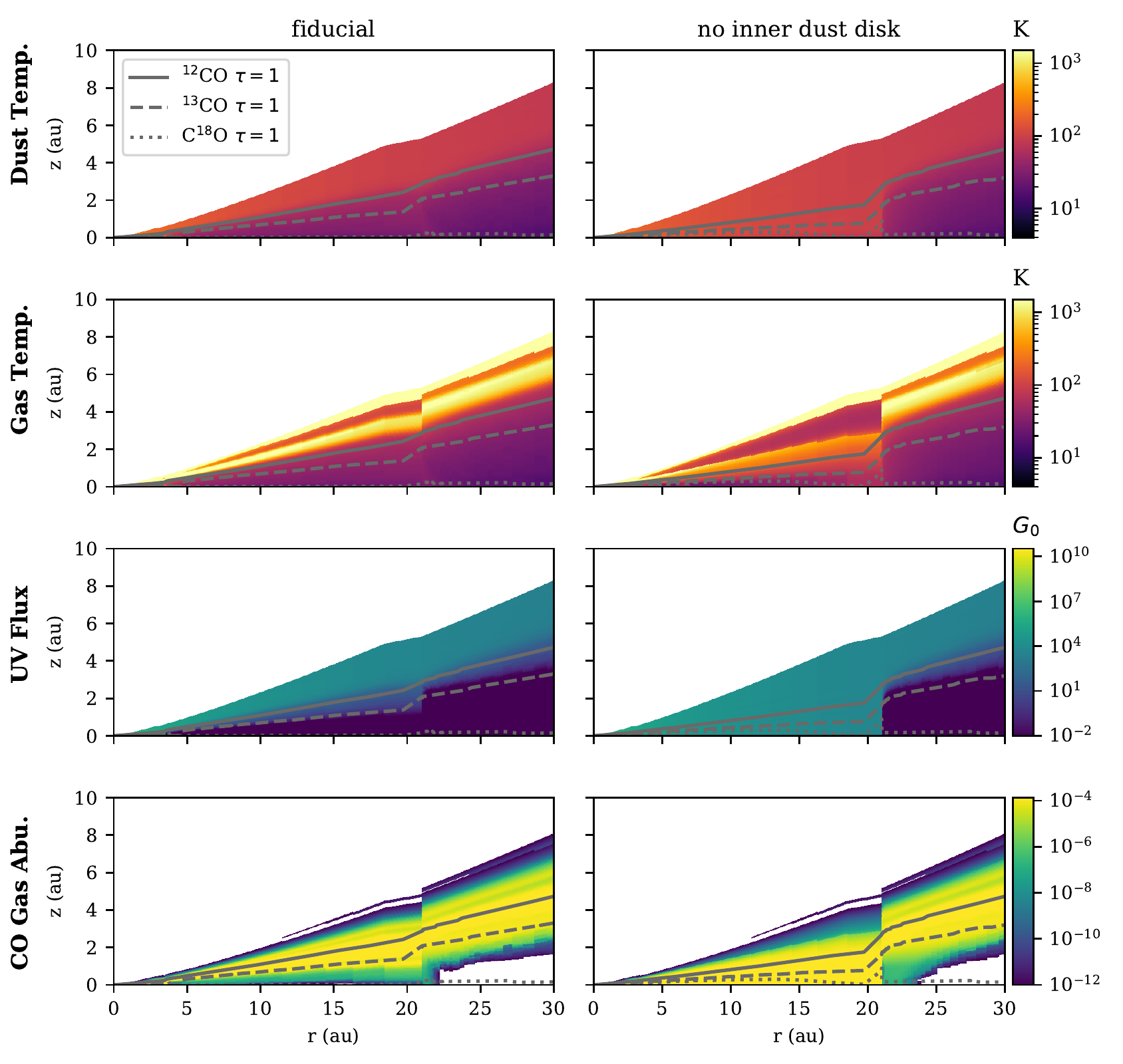}
    \includegraphics[scale=0.9]{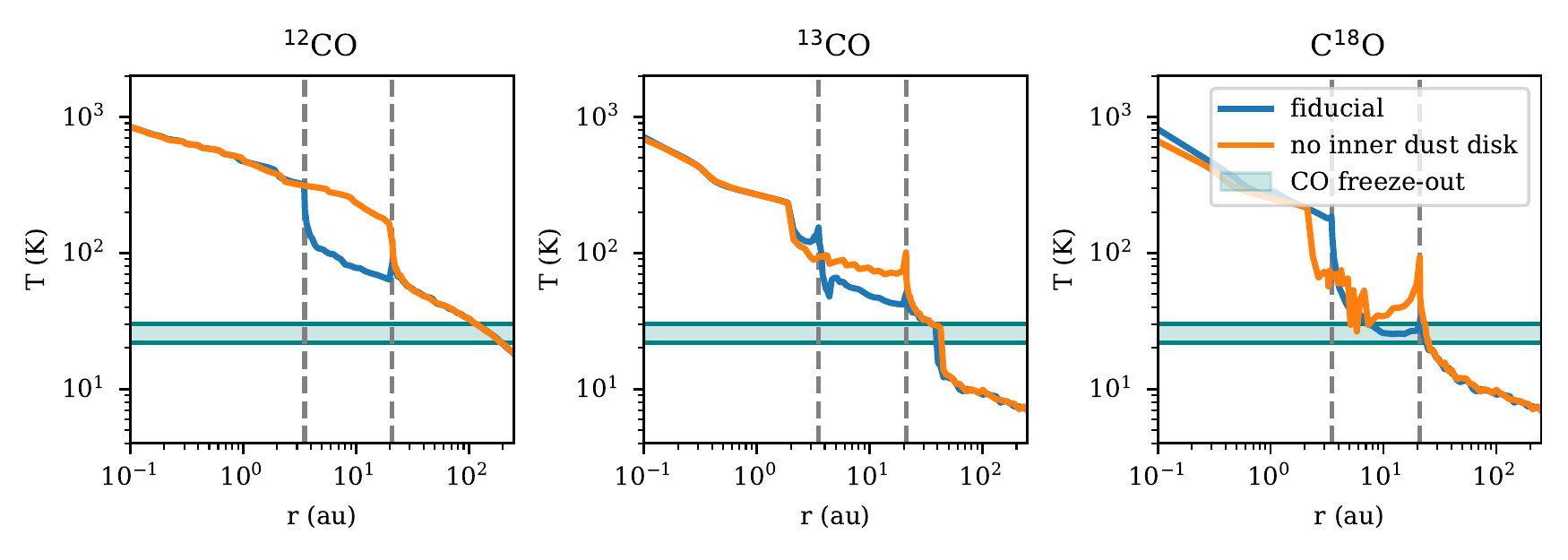}
    \caption{Comparison of selected 2D output variables for the DALI fiducial model (left column) and the same DALI model with no inner dust disk within 30 au (right column), as described in Section \ref{ssec:inner_disk}. First row: dust temperature; second row: gas temperature; third row CO gas abundance; fourth row: UV Flux in ISM units. Bottom row: Comparison of the gas temperature at the $\tau=1$ surface for the $^{12}$CO, $^{13}$CO, and C$^{18}$O lines, the blue shaded region indicates the CO freezeout temperature range. The $\tau=1$ surface for the $^{12}$CO, $^{13}$CO, and C$^{18}$O is overlaid as solid, dashed, and dotted lines respectively. }
    \label{fig:inner_disk_comp_dali}
\end{figure*}

\begin{figure*}[htb]
    \centering
    \includegraphics[]{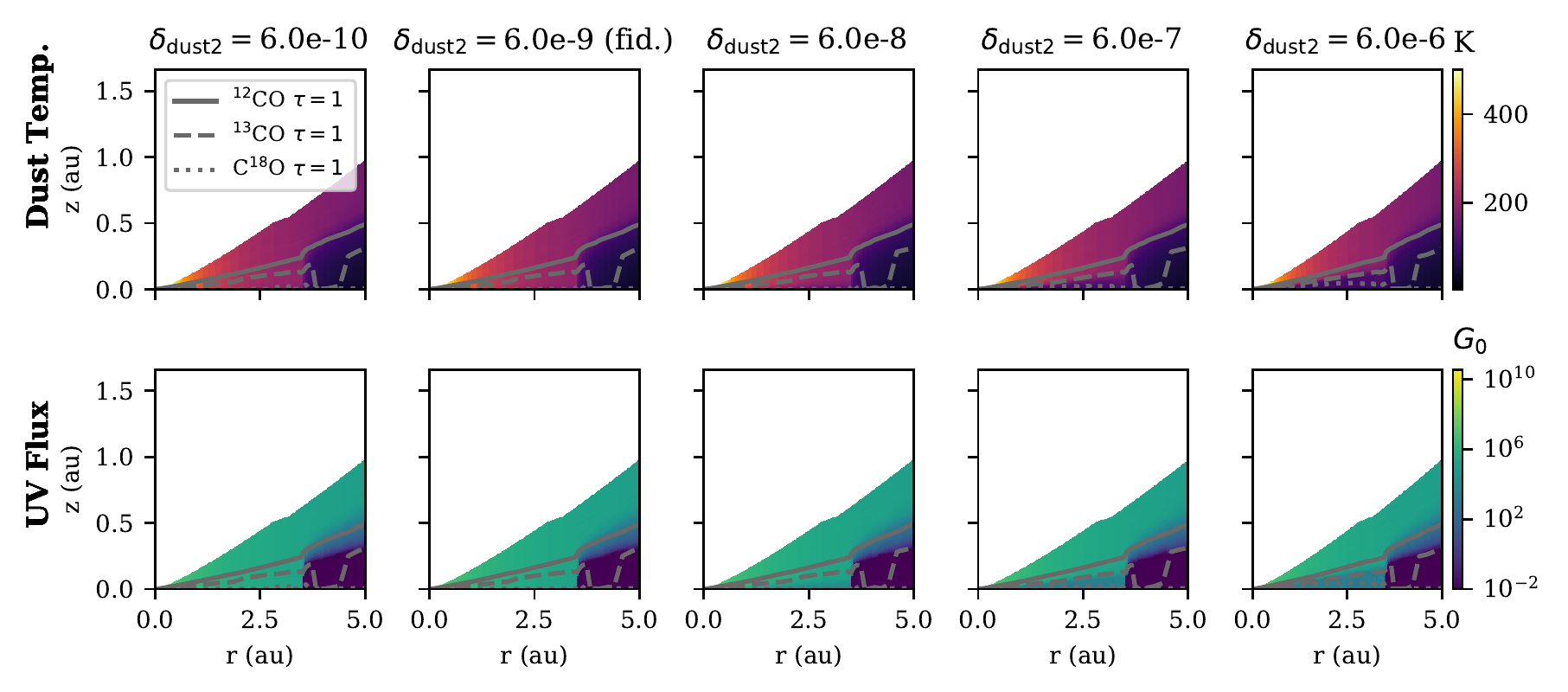}
    \caption{As Figure \ref{fig:inner_disk_comp_dali}, but for the comparison of different values of $\delta_\text{dust2}$ within 5 au, as described in Section \ref{ssec:dust_surf}. First row: dust temperature; second row: UV Flux in ISM units.}
    \label{fig:ddust2_comp_dali}
\end{figure*}

\begin{figure*}[htb]
    \centering
    \includegraphics[]{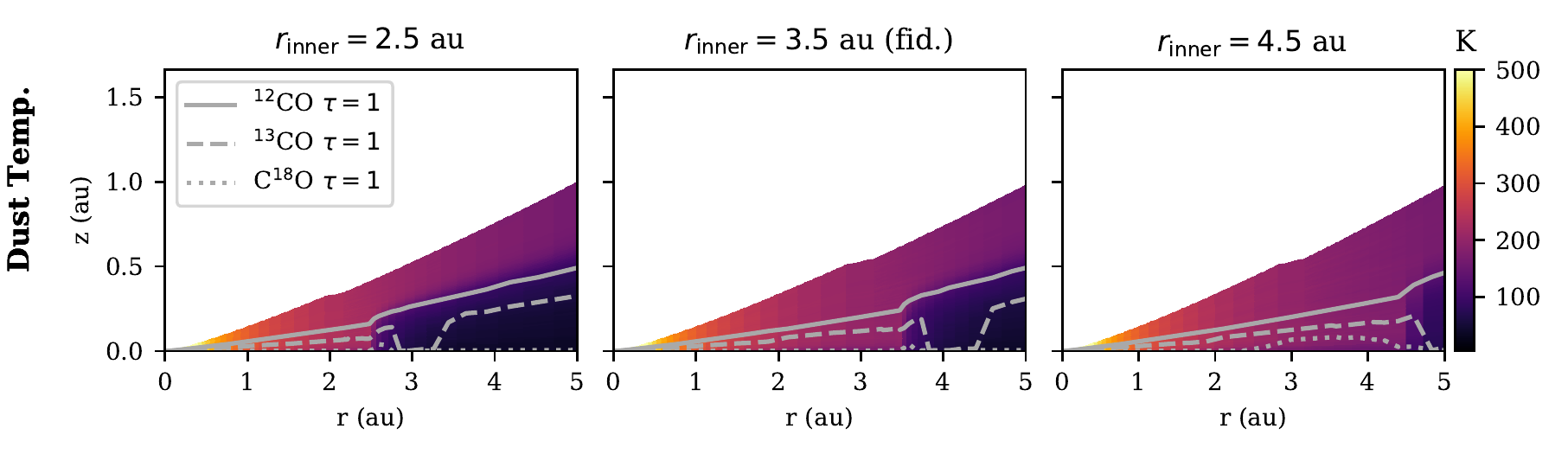}
    \caption{As Figure \ref{fig:inner_disk_comp_dali}, but for comparison of the effect on the dust temperature for different values of $r_\text{inner}$ within 5 au (right column), as described in Section \ref{ssec:dust_surf}.}
    \label{fig:rinner_comp_dali}
\end{figure*}

\begin{figure*}[htb]
    \centering
    \includegraphics[]{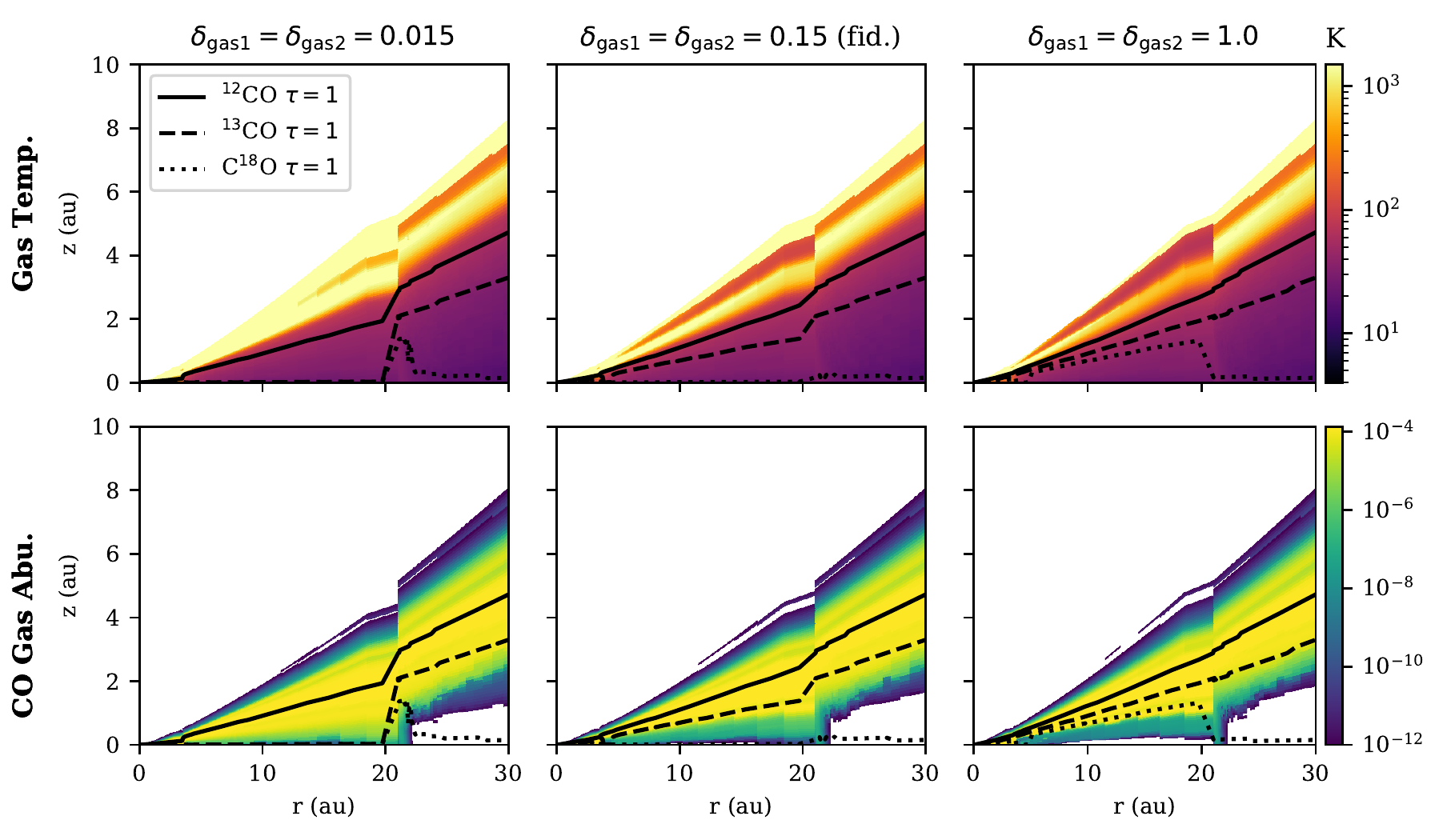}
    \includegraphics[]{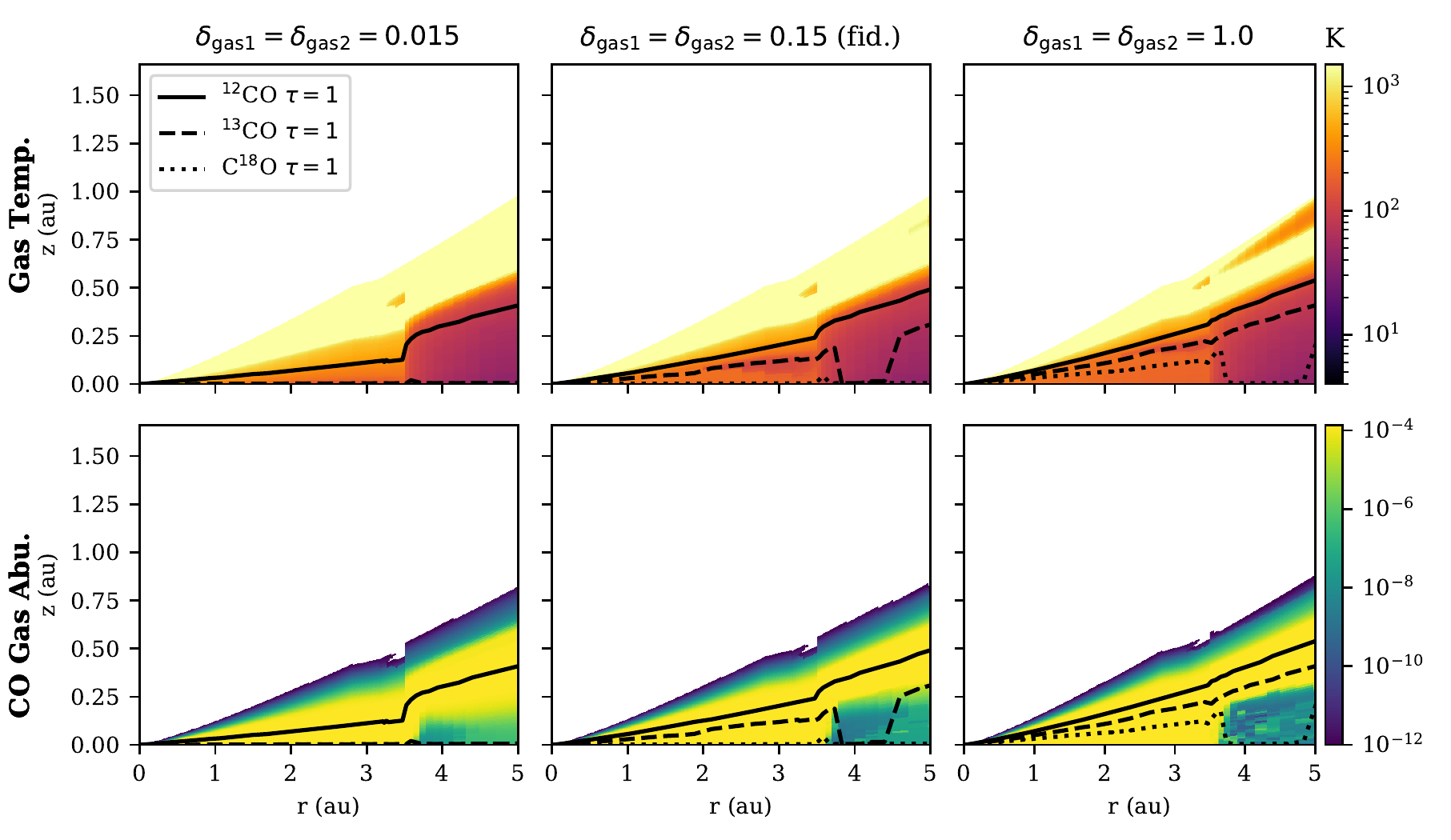}
    \caption{As Figure \ref{fig:inner_disk_comp_dali}, but for the comparison of different values of $\delta_\text{gas1}$ and $\delta_\text{gas2}$, as described in Section \ref{ssec:gas_comp}. First row: gas temperature within 30 au; second row: CO gas abundance within 30 au; third row: gas temperature within 5 au; fourth row: CO gas abundance within 5 au.}
    \label{fig:delta_gas_12_comp_dali}
\end{figure*}

\begin{figure*}[htb]
    \centering
    \includegraphics[]{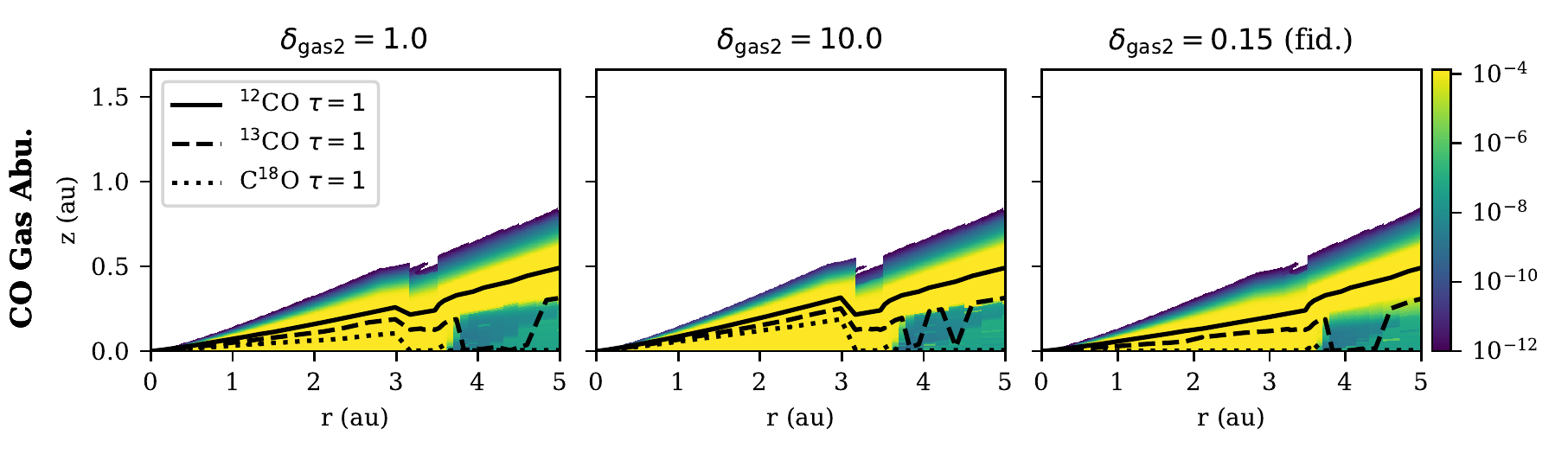}
    \caption{As Figure \ref{fig:inner_disk_comp_dali}, but for the comparison of the effect on the CO gas abundance of different values of $\delta_\text{gas2}$ within 5 au, as described in Section \ref{ssec:gas_comp}.}
    \label{fig:delta_gas_2_comp_dali}
\end{figure*}

\begin{figure}[htb]
    \centering
    \includegraphics[]{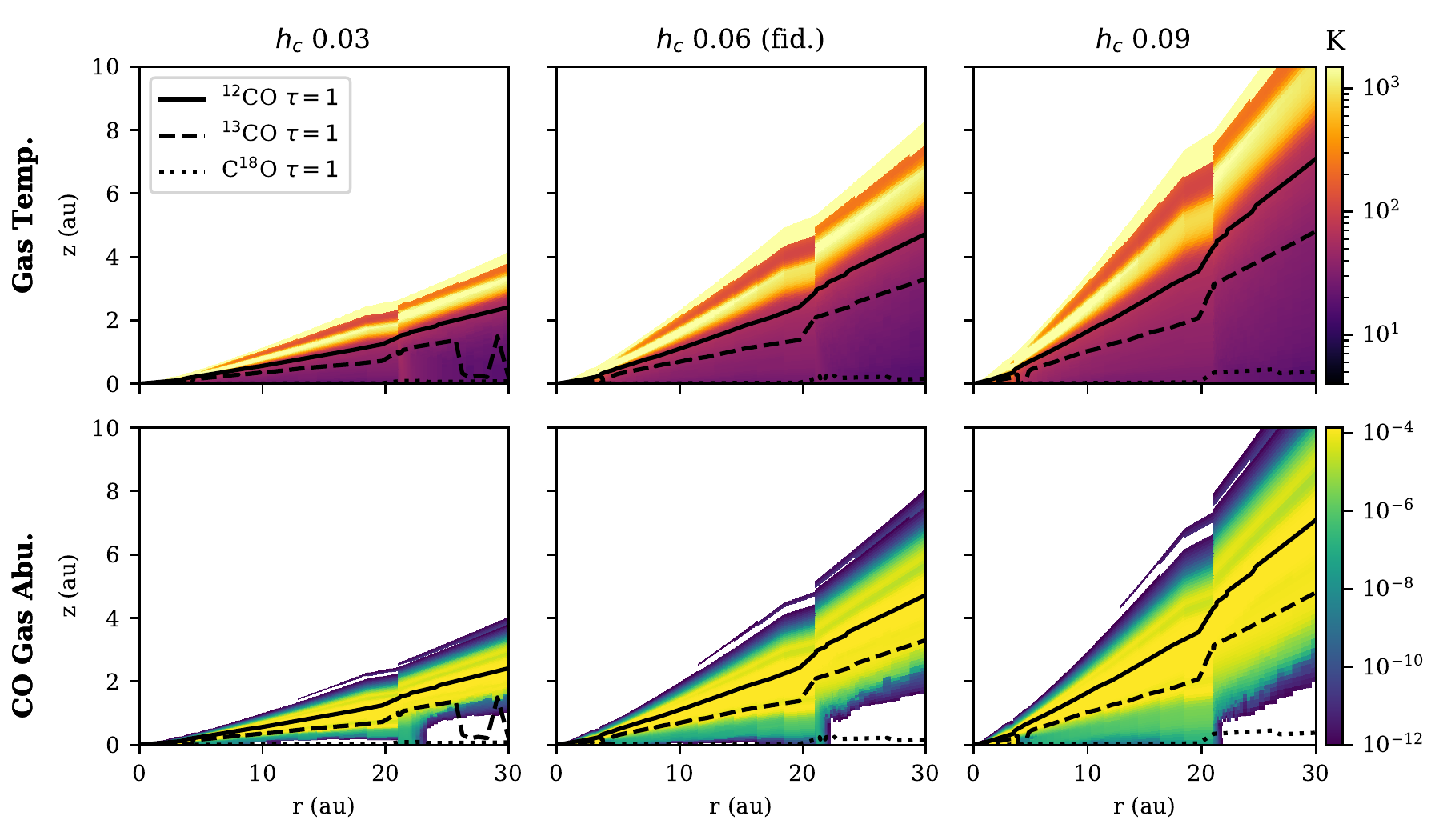}
    \includegraphics{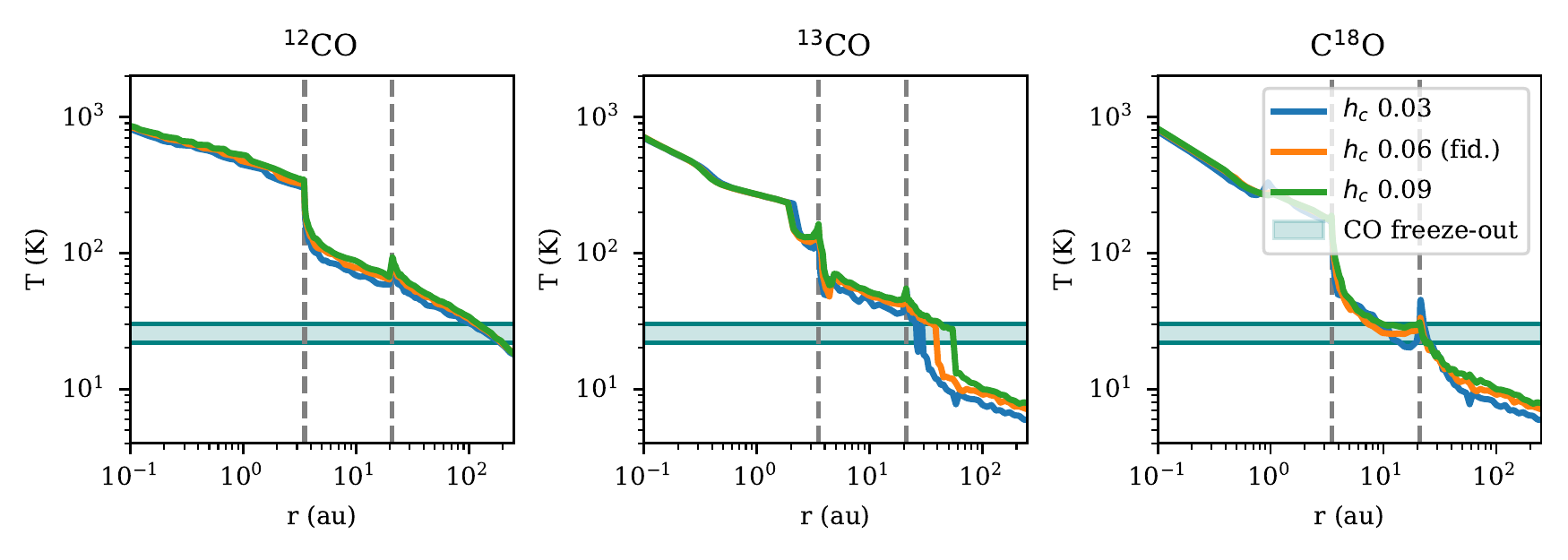}
    \caption{As Figure \ref{fig:inner_disk_comp_dali}, but for the comparison of different values of $h_c$ within 30 au, as described in Section \ref{ssec:scale_height}. First row: gas temperature; second row: CO gas abundance; third row: Comparison of the gas temperature at the $\tau=1$ surface for the $^{12}$CO, $^{13}$CO, and C$^{18}$O lines.}
    \label{fig:h_c_comp_dali}
\end{figure}

\begin{figure}[htb]
    \centering
    \includegraphics[]{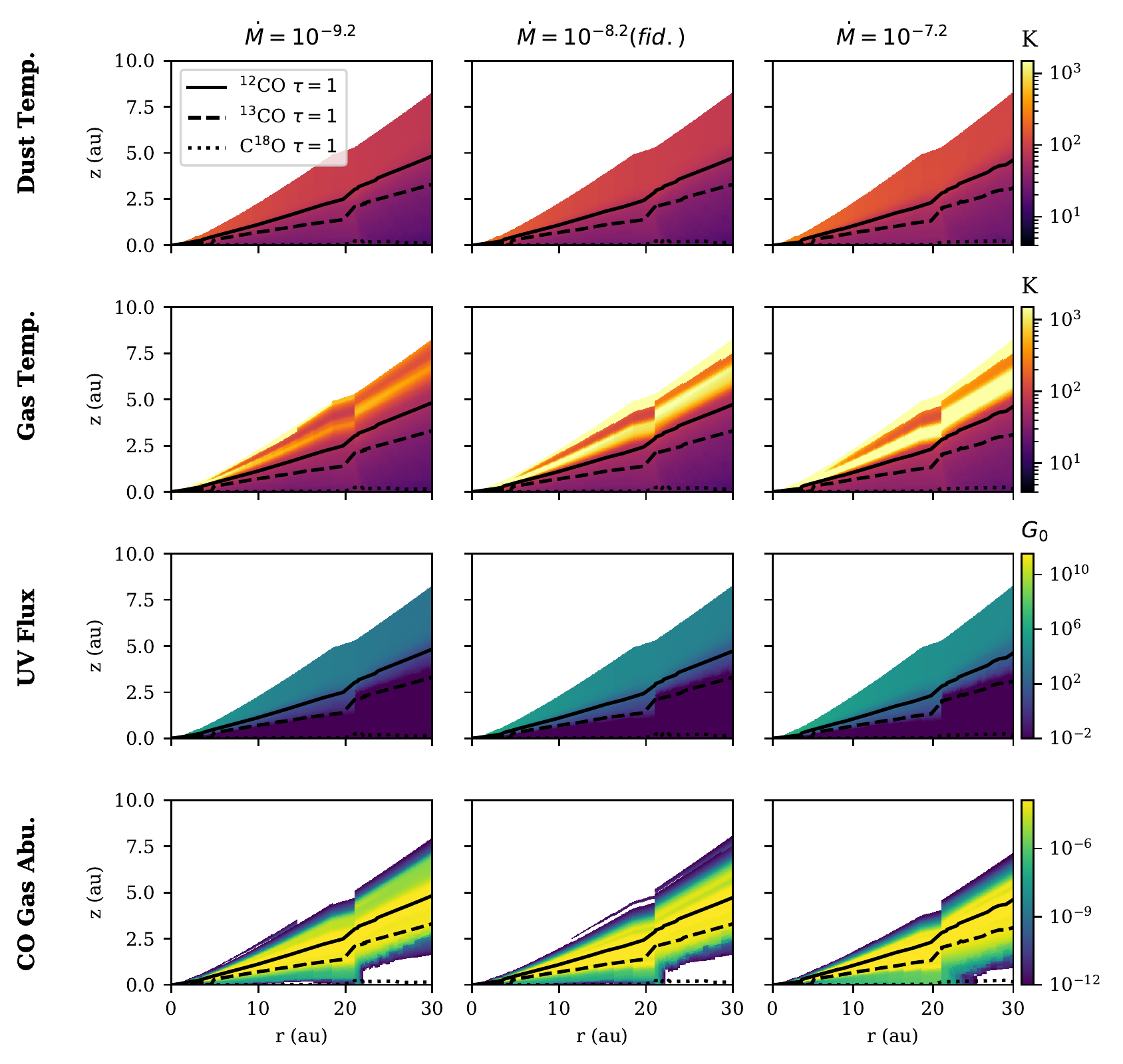}
    \caption{As Figure \ref{fig:inner_disk_comp_dali}, but for the comparison of different values of the stellar accretion rate within 30 au, as described in Section \ref{ssec:mdot}. First row: dust temperature within; second row: gas temperature; third row CO gas abundance; fourth row: UV flux in ISM units.}
    \label{fig:mdot_comp_dali}
\end{figure}

\begin{figure}[htb]
    \centering
    \includegraphics[]{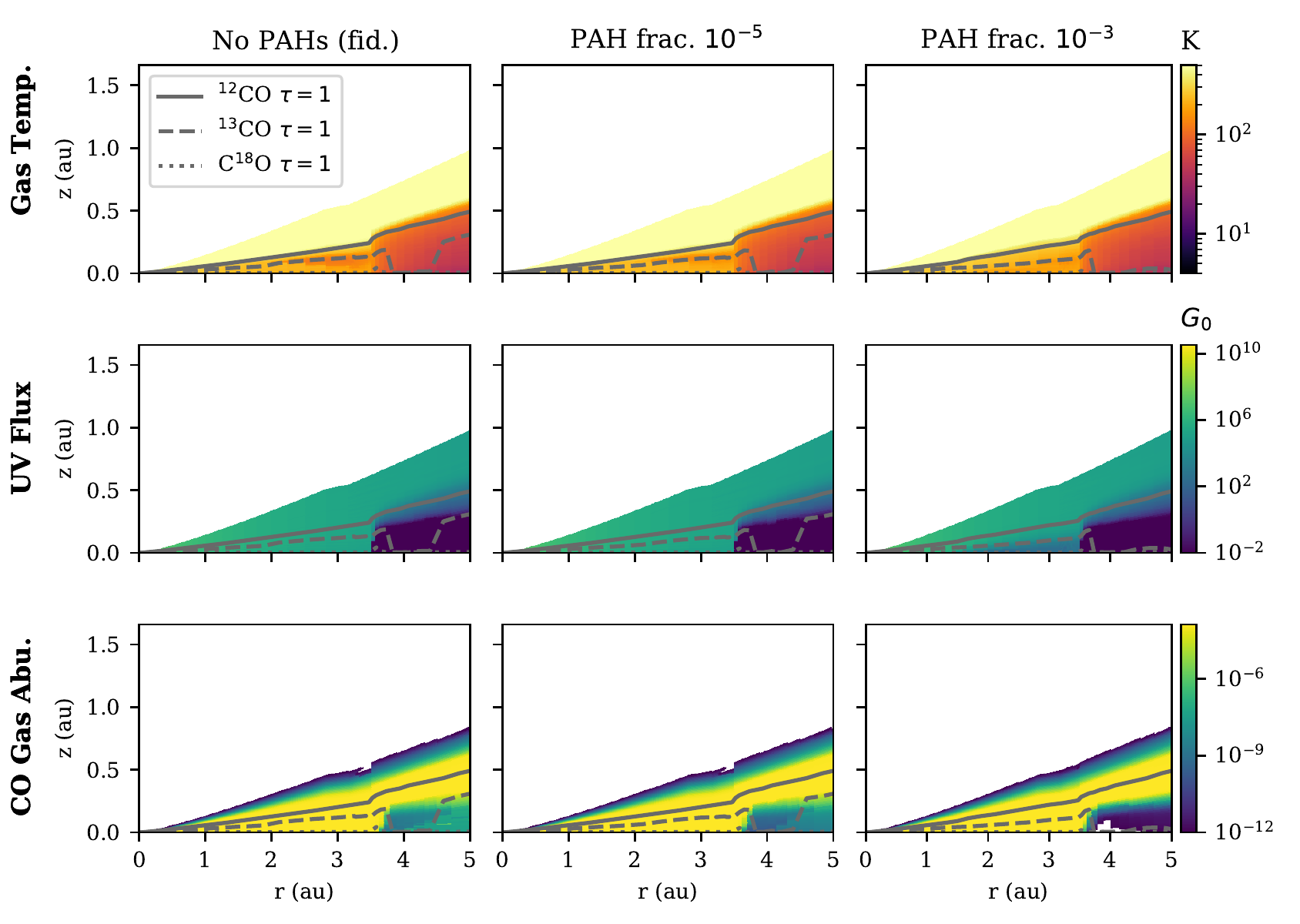}
    \caption{As Figure \ref{fig:inner_disk_comp_dali}, but for the comparison of different values of the global PAH abundance within 5 au, as described in Section \ref{ssec:pahs}. First row: UV flux in ISM units; second row: gas temperature; third row CO gas abundance.}
    \label{fig:pah_comp_dali}
\end{figure}

In this section, Figures \ref{fig:inner_disk_comp_dali} to \ref{fig:pah_comp_dali} show comparisons of physical quantities between sets of DALI models for the parameter space explored in Section \ref{ssec:parameter_space}.

\bibliography{bibliography}{}
\bibliographystyle{aasjournal}



\end{document}